\documentclass[acmsmall,screen,nonacm]{acmart}

\AtBeginDocument{%
  }

\usepackage{listings}
\usepackage{boogie}
\usepackage{silver}
\usepackage{color}
\usepackage{tikz}
\usetikzlibrary{positioning}
\usepackage{stmaryrd}
\usepackage{prooftree}
\usepackage[b]{esvect}
\usepackage{ifthen}

\def\istr{} 
\newcommand\appreftr[2]{
    \ifx\istr\undefined%
        App.~#2{ of the TR~\cite{TechnicalReport}}%
    \else
        App.~\ref{#1}%
    \fi}

\usepackage[scaled=0.9]{beramono}
\usepackage[T1]{fontenc}

\usepackage[normalem]{ulem}

 \def\nocolour{ }

\newcommand{\soutifcolour}[1]{\ifdefined\nocolour{}\else{\protect\sout{#1}}\fi}

\newcommand\todo[1]{\ifdefined\nocolour{}\else{\textcolor{red}{TODO: #1}}\fi}

\definecolor{gcolor}{rgb}{0.55, 0.71, 0.0}
\newcommand{\gaurav}[1]{\ifdefined\nocolour{#1}\else{\color{gcolor}{#1}}\fi}
\newcommand{\gfootnote}[1]{\ifdefined\nocolour{}\else\gaurav{\footnote{\gaurav{GAURAV: #1}}}\fi}
\newcommand{\gout}[1]{\gaurav{{\soutifcolour{#1}}}}

\newcommand\code[1]{\texttt{#1}}
\newcommand\secref[1]{Sec.~\ref{#1}}

\newcommand\tabref[1]{Tab.~\ref{#1}}
\newcommand\figref[1]{Fig.~\ref{#1}}

\newcommand{\wrt}{{{w.r.t.\@}}}
\newcommand{\eg}{{{e.g.\@}}}
\newcommand{\ie}{{{i.e.\@}}}

\newcommand{\eqdef}{\triangleq}
\newcommand{\translatesto}{\ensuremath{\leadsto}}

\newcommand{\vexhaleAux}[1]{\vkeyword{remcheck}\;#1}
\newcommand{\vexhaleAuxNoArg}[1]{\vkeyword{remcheck}}

\newcommand{\correctProcBpl}[2]{\ensuremath{\textsf{Correct}^{#1}_b(#2)}}
\newcommand{\correctMethodVpr}[3]{\ensuremath{\textsf{Correct}^{#1,#2}_v(#3)}}
\newcommand{\relMethodProc}[5]{\ensuremath{\textsf{Rel}_{#1,#2}^{#3}(#4,#5)}}
\newcommand{\specWellDefVpr}[1]{\ensuremath{\textsf{SpecWf}(#1)}}

\newcommand{\declsWellDefBpl}[4]{\textsf{DeclsWf}_{#1,#2}(#3,#4)}
\newcommand{\axiomsSatBpl}[4]{\textsf{AxiomSat}_#1(#2,#3,#4)}

\newcommand{\normalState}[1]{\textsf{N}(#1)}
\newcommand{\magicState}{\textsf{M}}
\newcommand{\failureState}[1]{\textsf{F}}

\newcommand{\ctxtVpr}{\ensuremath{\Gamma_v}}
\newcommand{\stateVpr}{\ensuremath{\sigma_v}}
\newcommand{\stateExtVpr}{\ensuremath{r_v}}

\newcommand{\leqVpr}{\ensuremath{\preceq}}

\newcommand{\mask}[1]{\ensuremath{\pi(#1)}}
\newcommand{\heap}[1]{\ensuremath{h(#1)}}
\newcommand{\store}[1]{\ensuremath{\textsf{st}(#1)}}

\newcommand{\normalVal}[1]{\textsf{V}(#1)}
\newcommand{\failureVal}{\lightning{}}

\newcommand{\redExprVpr}[3]{\ensuremath{\langle #1,#2 \rangle \Downarrow #3}}
\newcommand{\redExprsVpr}[3]{\ensuremath{\langle #1,#2 \rangle [\Downarrow] #3}}

\newcommand{\redStmtVpr}[4]{\ensuremath{#1 \vdash \langle #2,#3 \rangle \rightarrow_\mathsf{v} #4}}
\newcommand{\redInh}[3]{\ensuremath{\langle #1,#2 \rangle \rightarrow_{\mathsf{inh}} #3}}
\newcommand{\redExhAux}[4]{\ensuremath{#1 \vdash \langle #2,#3 \rangle \redExhSym #4}}
\newcommand{\redExhSym}{\rightarrow_{\mathsf{rc}}}
\newcommand{\stmtExhSuccSemRuleName}{\rulename{exh-succ}}
\newcommand{\stmtExhFailSemRuleName}{\rulename{exh-fail}}

\newcommand{\exhAuxSepSemRuleName}{\rulename{rc-sep}}
\newcommand{\exhAuxAccSemRuleName}{\rulename{rc-acc}}

\newcommand{\initCtxtVpr}[3]{\textsf{initCtxt}_v^{#1,#2}(#3)}

\newcommand{\precondVpr}[1]{\textsf{pre}(#1)}
\newcommand{\postcondVpr}[1]{\textsf{post}(#1)}
\newcommand{\bodyVpr}[1]{\textsf{body}(#1)}

\newcommand{\exhNonDetSelectNoArg}{\textsf{nonDet}}
\newcommand{\exhNonDetSelect}[3]{\exhNonDetSelectNoArg{}(#1,#2,#3)}
\newcommand{\removePerm}[4]{\textsf{rem}(#1,#2,#3,#4)}

\newcommand{\exhAuxExprEvalState}{expression evaluation state}
\newcommand{\exhAuxReductionState}{reduction state}

\newcommand{\ctxtBpl}{\ensuremath{\Gamma_b}}
\newcommand{\stateBpl}{\ensuremath{\sigma_b}}
\newcommand{\stateExtBpl}{\ensuremath{r_b}}
\newcommand{\progPointBpl}{\ensuremath{\gamma}}

\newcommand{\redStmtBpl}[5]{\ensuremath{#1 \vdash (#2,#3) \rightarrow_{\mathsf{b}}^* (#4,#5)}}

\newcommand{\initCtxtBpl}[4]{\ensuremath{\textsf{initCtxt}_b}^{#1}(#2,#3,#4)}
\newcommand{\initProgPoint}[1]{\textsf{init}_b(#1)}

\newcommand{\genericSimNoArg}{\textsf{sim}}
\newcommand{\genericSim}[7]{\genericSimNoArg{}_{#1}(#2,#3,#4,#5,#6,#7)}

\newcommand{\stmtSimNoArg}{\textsf{stmSim}}
\newcommand{\stmtSim}[7]{\stmtSimNoArg{}_{#1,#2}(#3,#4,#5,#6,#7)}

\newcommand{\exhAuxSimNoArg}{\textsf{rcSim}}
\newcommand{\exhAuxSim}[6]{\exhAuxSimNoArg{}_{#1}(#2,#3,#4,#5,#6)}

\newcommand{\exhAuxInvSimNoArg}{\textsf{rcInvSim}}
\newcommand{\exhAuxInvSim}[7]{\exhAuxInvSimNoArg{}_{#1}^{#2}(#3,#4,#5,#6,#7)}

\newcommand{\exprWfSimNoArg}{\textsf{wfSim}}
\newcommand{\exprWfSim}[6]{\exprWfSimNoArg{}_{#1}(#2,#3,#4,#5,#6)}

\newcommand{\exprsWfSimNoArg}{\textsf{wfSim}}
\newcommand{\exprsWfSim}[6]{\exprsWfSimNoArg{}_{#1}(#2,#3,#4,#5,#6)}

\newcommand{\onlyBoogieSimNoArg}{\textsf{bSim}}
\newcommand{\onlyBoogieSim}[5]{\onlyBoogieSimNoArg{}_{#1}(#2,#3,#4,#5)}

\newcommand{\compRuleName}{\rulename{comp}}
\newcommand{\propRuleName}{\rulename{bprop}}
\newcommand{\exhAuxPropRuleName}{\rulename{rcprop}}
\newcommand{\conseqRuleName}{\rulename{cons}}

\newcommand{\stmtSeqRuleName}{\rulename{seq-sim}}
\newcommand{\stmtSimExhRuleName}{\rulename{exh-sim}}

\newcommand{\exhAuxInvSepRuleName}{\rulename{rsep-sim}}
\newcommand{\exhAuxNoInvSepRuleName}{\rulename{rsep2-sim}}
\newcommand{\exhAuxAccRuleName}{\rulename{racc-sim}}

\newcommand{\stateRelInstName}{\textsf{SR}}
\newcommand{\stateRelInstBase}[3]{\stateRelInstName_{#1}^{#2,#3}}
\newcommand{\stateRelInst}[6]{\stateRelInstBase{#1}{#2}{#3}((#4,#5),#6)}

\newcommand{\trRecord}{\ensuremath{\mathit{Tr}}}
\newcommand{\varTr}[1]{\ensuremath{\mathit{var}(#1)}}
\newcommand{\fieldTr}[1]{\ensuremath{\mathit{field}(#1)}}

\newcommand{\heapTr}[1]{\ensuremath{\mathit{H}(#1)}}
\newcommand{\maskTr}[1]{\ensuremath{\mathit{M}(#1)}}

\newcommand{\heapDefTr}[1]{\ensuremath{\mathit{H^0}(#1)}}
\newcommand{\maskDefTrName}{\ensuremath{\mathit{M^0}}}
\newcommand{\maskDefTr}[1]{\ensuremath{\maskDefTrName(#1)}}
\newcommand{\auxVars}{\ensuremath{\mathit{AV}}}
\newcommand{\storeRelNoArg}{\textsf{stRel}}
\newcommand{\storeRel}[4]{\storeRelNoArg{}_{#1}(#2,#3,#4)}
\newcommand{\heapMaskRelNoArg}{\textsf{hmRel}}
\newcommand{\heapMaskRel}[5]{\heapMaskRelNoArg{}_{#1}(#2,#3,#4,#5)}
\newcommand{\fieldRelNoArg}{\textsf{fieldRel}}
\newcommand{\fieldRel}[3]{\fieldRelNoArg{}_{#1}(#2,#3)}

\newcommand{\consistent}[1]{\textsf{consistent}(#1)}

\definecolor{CertLinkColor}{rgb}{0.0,0.3,0.55}
\definecolor{FinalCertLinkColor}{rgb}{0.0,0.7,0.0}

\newcommand{\wfAccAssmsNoArg}{\textsf{wfAccSucc}}
\newcommand{\wfAccAssms}[5]{\wfAccAssmsNoArg(#1,#2,#3,#4,#5)}
\definecolor{wfAccAssmsColor}{rgb}{0.58, 0.0, 0.83}

\newcommand{\exhAccSuccNoArg}{\textsf{exhAccSucc}}
\newcommand{\exhAccSucc}[3]{\exhAccSuccNoArg(#1,#2,#3)}
\definecolor{exhAccSuccColor}{rgb}{0.71, 0.4, 0.11}

\newcommand{\auxVarMap}{auxiliary variable map}

\makeatletter

\newskip \point

\def \premisespacing{\quad}

\def \RulePremisesNewlineMore[#1]#2.#3#4{\@ifnextchar\bgroup{\RulePremisesNewlineMore[#1]{#2}.{#3\premisespacing#4}}{\@ifnextchar.{\RulePremisesNewline[#1]{{\begin{array}{c}#2\\#3\premisespacing#4\end{array}}}}{\RuleMultiPremise[#1]{{\begin{array}{c}#2\\#3\end{array}}}{#4}}}}

\def \RulePremisesNewline[#1]#2.#3{\@ifnextchar\bgroup{\RulePremisesNewlineMore[#1]{#2}.{#3}}{\@ifnextchar.{\RulePremisesNewline[#1]{{\begin{array}{c}#2\\#3\end{array}}}}{\RuleMultiPremise[#1]{#2}{#3}}}}

\def \RuleMultiPremise[#1]#2#3{\@ifnextchar\bgroup{\RuleMultiPremise[#1]{#2\premisespacing#3}}{\@ifnextchar.{\RulePremisesNewline[#1]{#2\premisespacing#3}}{\prooftree #2\justifies#3 \using{#1}\endprooftree}}}

\def \RuleWithName[#1]#2{\@ifnextchar\bgroup {\RuleMultiPremise[#1]{#2}}{\@ifnextchar.{\RulePremisesNewline[#1]{#2}}{\prooftree \justifies #2 \using{#1} \endprooftree}}}

\def \RuleWithInfo[#1]{\@ifnextchar[{\RuleWithNameAndCondition[#1]}{\RuleWithName[(#1)]}}

\def \RuleWithNameAndCondition[#1][#2]{\RuleWithName[(#1)^{#2}]}

\def \Inf{\proofrulebaseline=2ex \abovedisplayskip12\point\belowdisplayskip12\point \abovedisplayshortskip8\point\belowdisplayshortskip8\point \@ifnextchar[{\RuleWithInfo}{\RuleWithName[ ]}}

\makeatother

\newcommand{\rulename}[1]{\textsc{#1}}

\begin{document}

\title[Formally Validating Translations into an Intermediate Verification Language]{Towards Trustworthy Automated Program Verifiers:\\ 
Formally Validating Translations into an Intermediate Verification Language (extended version)}

\author{Gaurav Parthasarathy}
\orcid{0000-0002-1816-9256}
\affiliation{%
  \institution{Department of Computer Science, ETH Zurich}
  \city{Zurich}
  \country{Switzerland}
}
\email{gaurav.parthasarathy@inf.ethz.ch}

\author{Thibault Dardinier}
\orcid{0000-0003-2719-4856}
\affiliation{%
  \institution{Department of Computer Science, ETH Zurich}
  \city{Zurich}
  \country{Switzerland}
}
\email{thibault.dardinier@inf.ethz.ch}

\author{Benjamin Bonneau}
\orcid{0009-0005-9688-1299}
\affiliation{%
  \institution{Universit\'e Grenoble Alpes - CNRS - Grenoble INP - VERIMAG}
  \city{Grenoble}
  \country{France}
}
\email{Benjamin.Bonneau@univ-grenoble-alpes.fr}

\author{Peter Müller}
\orcid{0000-0001-7001-2566}
\affiliation{%
  \institution{Department of Computer Science, ETH Zurich}
  \city{Zurich}
  \country{Switzerland}
}
\email{peter.mueller@inf.ethz.ch}

\author{Alexander J. Summers}
\orcid{0000-0001-5554-9381}
\affiliation{%
  \institution{University of British Columbia}
  \city{Vancouver}
  \country{Canada}
}
\email{alex.summers@ubc.ca}


\begin{abstract}

Automated program verifiers are typically implemented using an intermediate verification language (IVL), such as Boogie or Why3. A verifier front-end translates the input program and specification into an IVL program, while the back-end generates proof obligations for the IVL program and employs an SMT solver to discharge them. Soundness of such verifiers therefore requires that the front-end translation faithfully captures the semantics of the input program and specification in the IVL program, and that the back-end reports success only if the IVL program is actually correct. For a verification tool to be trustworthy, these soundness conditions must be satisfied by its  \emph{actual implementation}, not just the program logic it uses. 

In this paper, we present a novel validation methodology that, given a formal semantics for the input language and IVL, provides formal soundness guarantees for front-end implementations. For each run of the verifier, we automatically generate a proof in Isabelle showing that the correctness of the produced IVL program implies the correctness of the input program. This proof can be checked independently from the verifier, in Isabelle, and can be combined with existing work on validating back-ends to obtain an end-to-end soundness result. Our methodology based on forward simulation employs several modularisation strategies to handle the large semantic gap between the input language and the IVL, as well as the intricacies  of practical, optimised translations. We present our methodology for the widely-used Viper and Boogie languages. Our evaluation shows that it is effective in validating the translations performed by the existing Viper implementation. 

\end{abstract}

\setcopyright{rightsretained}
\copyrightyear{2024}
\acmJournal{PACMPL}
\acmYear{2024} 
\acmVolume{8} 
\acmNumber{PLDI} 
\acmArticle{208} 
\acmMonth{6}
\acmDOI{10.1145/3656438}




\maketitle

\section{Introduction} 
Program verifiers are tools that try to automatically establish the correctness of an input program with respect to a specification. A standard approach for achieving automation is to reduce the input program and specification to a set of first-order formulas whose validity implies the correctness of the input program; the validity of formulas is automatically checked using an SMT solver. Instead of directly producing logical formulas, many program verifiers are \emph{translational verifiers}: they translate an input program and specification into a program in an \emph{intermediate verification language (IVL)}; we call this a \emph{front-end translation}. An IVL comes with its own \emph{back-end verifier} that ultimately reduces IVL programs to logical formulas. This translational approach via an IVL allows for the reuse of the IVL's back-end technology across multiple front-end verifiers, and makes for a more understandable target representation than direct mappings to logical formulas, simplifying the development of state-of-the-art program verifiers.
%

A very wide variety of practical program verifiers are translational verifiers; \eg{}
Corral~\cite{LalQ14}, 
Dafny~\cite{LeinoDafny10},
SMACK~\cite{CarterHWRE16},   
SYMDIFF~\cite{LahiriHKR12},  
and Viper~\cite{MuellerSchwerhoffSummers16}
target the imperative Boogie IVL~\cite{LeinoBoogie2}, while Creusot~\cite{DenisJM22} and Frama-C~\cite{framac15} translate to the functional Why3 IVL~\cite{why3}. Multiple layers of front-end translations and IVLs can also be \emph{composed} (\eg{} Prusti~\cite{AstrauskasMuellerPoliSummers19b} builds on Viper as an IVL).

To ensure that successful verification indeed implies that the input program satisfies its specification, any translational verifier must meet two \emph{soundness conditions}: (1)~\emph{Front-end soundness}: the \emph{translation} into the IVL is faithful, \ie{} correctness of the produced IVL program implies correctness of the input program, and (2)~\emph{IVL back-end soundness}: if the back-end IVL verifier reports success, the IVL program is correct. Trustworthiness of program verifiers requires formal guarantees for both soundness conditions. It is \emph{not} sufficient to prove soundness of the program logics they employ in principle: automated verifiers are complex systems, and it is essential that formal guarantees also cover their \emph{actual implementations}, where soundness bugs can and do arise. 

Existing work on ensuring front-end soundness is based on idealised implementations that are formalised on paper or in an interactive theorem prover. In practice, practical front-end translations are implemented in efficient mainstream programming languages, use diverse libraries and programming paradigms, and include subtle optimisations omitted from idealised implementations; there is a very large gap between the translations proved correct and the actual translations used in practice. In this paper, we bridge this gap for the first time, developing an approach to formally validate the front-end soundness of translations used in \emph{existing, practical} verifier implementations. IVL back-end verifier soundness, which includes the soundness of the underlying SMT solver, is a better-studied and orthogonal concern; our results can be combined with work in that area to obtain end-to-end guarantees for an entire verification toolchain~\cite{BohmeW10,EkiciMTKKRB17,verit-hol-2019,ParthasarathyMuellerSummers21,Garchery21}.

Proving front-end soundness once and for all for a realistic verifier implementation is practically infeasible, since such implementations are large (e.g.\ 17.2 KLOC and 8.5 KLOC for the Dafny-to-Boogie and Viper-to-Boogie front-ends, respectively) and are typically written in languages that lack a full formalisation (C{\#} and Scala, in the examples above). 
Instead, we develop a translation validation approach that, given a formal semantics for the input language and IVL, \emph{automatically} generates a formal proof on every run of the verifier via an instrumentation of the existing implementation.
Our proofs are expressed in the Isabelle theorem prover~\cite{NipkowPW02}, and thus can be checked independently, effectively removing the (substantial) front-end translation from the trusted code base of the verifier.

\paragraph{Challenges.}

Formally validating front-end translations is challenging for three main reasons:

\textit{1.~Semantic gap:}
There is a large semantic gap between a front-end language and an IVL, which concerns 
the state model (\eg\ neither Boogie nor Why3 have a heap, but most front-end languages do),
the execution model (\eg{} Viper heap accesses are partial operations that must be guarded by semantic conditions ultimately checked by verification, while Boogie and Why3 use syntactic checks to guard state accesses such as disallowing global variables in Boogie axioms and restricting aliasing between mutable variables in Why3),
and the program logics used to reason about programs (\eg{} front-ends use complex logics, such as dynamic frames~\cite{Kassios06} in Dafny, a flavour of separation logic~\cite{SmansJacobsPiessens12,ParkinsonSummers12} in Viper, and prophetic reasoning in Creusot~\cite{DenisJM22}, whereas Boogie and Why3 do not have built-in support for such logics). 
To bridge the semantic gap, front-ends translate input programs into a complex combination of low-level operations and background logical axiomatisations of input language concepts; validation needs to precisely account for the combination of these ingredients, while allowing the separation of translation aspects for the sake of modularity and maintainability.

\textit{2.~Diverse translations:}
Practical front-end translations are \emph{diverse} in the sense that they use multiple alternative translations for the same feature, \eg{} more efficient translations that are sound only in certain cases. 
These translations also evolve frequently over time, as new techniques and features are developed or optimised; ideally a formal approach to validation should provide means of minimising the impact of the exchange of one translation for another.

\textit{3.~Non-locality:}
The soundness of the translation of a fragment of the input program may depend on several checks that are performed at different places in the IVL program. For instance, the translation of a procedure call might be sound only because well-formedness of the procedure specification has been checked elsewhere in the generated IVL code. Such non-local checks are commonly used to speed up verification, for instance, to check well-formedness conditions once and for all rather than each time a specification is used. However, they complicate the soundness argument, which needs to somehow track the dependencies on properties checked elsewhere.

\paragraph{This paper.}

We present the first approach for enabling automatic formal validation for existing implementations of the front-end translations employed in many practical program verifiers. This validation guarantees front-end soundness and, thus, makes automated program verifiers substantially more trustworthy. 

The core of our approach is a general methodology for generating \emph{forward simulations}~\cite{LynchV95} between the statements of the input and the IVL program in a modular way. Our methodology provides solutions to the three challenges above. It (1)~bridges the semantic gap with a novel approach by which the simulation proof is split into smaller simulations,
(2)~supports diverse translations by expressing simulations abstractly, and
(3)~handles non-locality by systematically and formally tracking dependencies  during a simulation proof.

For concreteness, we present our methodology for the translation from a core fragment of Viper to Boogie, as implemented in an existing and actively-used verification tool~\cite{CarbonGithub}. This translation is significant because it exhibits all of the challenges discussed above and because both Viper and Boogie are widely used. For instance, Viper is used in Gobra (Go)~\cite{WolfACOPM21}, Prusti (Rust)~\cite{AstrauskasMuellerPoliSummers19b}, Nagini (Python)~\cite{Eilers18}, VerCors (Java)~\cite{BlomHuisman14}, and 
Gradual C0~\cite{divincenzo2022gradual}. The  soundness of each of these tools relies on the Viper verifiers being sound. Note that these tools use Viper as an IVL, but for the purpose of this paper, we will treat it as a front-end language that is translated to Boogie. While our methodology is phrased in terms of Viper and Boogie, we have designed our approach, which solves the key challenges above, to generalise to other front-end translations (\eg{} the Dafny-to-Boogie translation).

\paragraph{Contributions.}

We make the following technical contributions:
\begin{itemize}
    \item We develop a general methodology for the automatic validation of front-end translations based on forward simulation proofs. We present this methodology for the translation from Viper to Boogie. As a foundation for the proofs, we formalise a semantics for a core subset of Viper in Isabelle and connect this with an existing Isabelle formalisation for Boogie~\cite{ParthasarathyMuellerSummers21}.

    \item We instrument the existing Viper-to-Boogie implementation such that, for a subset of Viper, it automatically generates an Isabelle proof justifying the soundness of the translation. These generated proofs can be checked independently in Isabelle, which ensures front-end soundness of the Viper verifier.

    \item Our evaluation on a diverse set of Viper programs demonstrates our approach's effectiveness: we were able to generate proofs and check them in Isabelle fully automatically in all cases.

    \item As part of justifying the axioms used in Boogie programs, we provide the first approach to formally deal with a restricted version of Boogie's (impredicatively-)\emph{polymorphic maps}~\cite{LeinoRPoly10}.
\end{itemize}

\paragraph{Outline.} \secref{sec:background} provides the necessary background on Viper and Boogie. \secref{sec:methodology} introduces our forward simulation methodology for relating Viper and Boogie statements.
\secref{sec:validation_impl} presents how we formally validate the existing Viper-to-Boogie implementation using our forward simulation methodology.
\secref{sec:evaluation} evaluates the proofs generated by our instrumentation. 
\secref{sec:related_work} presents related work and \secref{sec:conclusion} concludes.
Our publicly-available artifact~\cite{artifact} contains the Isabelle formalisation for \secref{sec:background}, \secref{sec:methodology}, and \secref{sec:validation_impl}, our proof-producing Viper-to-Boogie implementation, and the examples used for the evaluation. Further details are available \ifx\istr\undefined in our technical report~\cite{TechnicalReport} (hereafter, TR)\else in the appendix\fi.

\section{Viper and Boogie: Background and Semantics}\label{sec:background}
In this section, we present the necessary background on the Viper and Boogie languages.
We introduce our supported Viper subset and the corresponding Boogie subset targeted by the pre-existing Viper-to-Boogie implementation (\secref{subsec:viper_and_boogie_overview}),
give an overview of the semantics of Boogie (\secref{subsec:boogie_semantics}) and Viper (\secref{subsec:viper_semantics}), and finally show an example of the translation used by the Viper-to-Boogie implementation 
(\secref{subsec:vpr_to_bpl_example}).

\begin{figure}[t]
\small
\begin{align*}
\textit{VExpr} \ni e ::=&\; x \mid \mathit{lit} \mid \vfieldacc{e}{f} \mid \binaryop{e}{\mathit{bop}}{e} \mid \unaryop{\mathit{uop}}{e} \quad
\textit{VAssert} \ni A ::=\; e \mid \vaccpred{e}{f}{e} \mid \vsep{A}{A} \mid \vimp{e}{A} \mid \vcondassert{e}{A}{A} \\
\textit{VStmt} \ni s ::=&\; \vassign{x}{e} \mid \vfieldassign{e}{f}{v} \mid \vmcallassign{\vec{y}}{m}{\vec{x}} \mid \vmcall{m}{\vec{x}} \mid \vvardecl{x}{\tau} \mid \vinhale{A} \mid \vexhale{A} \mid \vassert{A} \mid \\
     &\; s;s \mid \vif{e}{s}{s}
\end{align*}
\begin{align*}
&\textit{BExpr} \ni e_b ::=\; x \mid \mathit{lit}_b \mid \binaryop{e_b}{\mathit{bop}}{e_b} \mid \unaryop{\mathit{uop}}{e_b} \mid \funcall{f}{\vec \tau_b}{\vec e_b} \mid \bforall{x}{\tau_b}{e_b} \mid \bexists{x}{\tau_b}{e_b} \mid \bforallt{t}{e_b} \mid \bexistst{t}{e_b} \\
&\textit{BSimpleCmd} \ni c_b ::=\; \bassume{e_b} \mid \bassert{e_b} \mid \bassign{x}{e_b} \mid \bhavoc{x} \quad\hspace{.2em}\quad \textit{BStmtBlock} \ni b_b ::=\; \bAstBlock{\overrightarrow{c_b}}{\textit{if}_b} \\
&\textit{BIfOpt} \ni \textit{if}_b ::=\; \bAstIf{e_b}{s_b}{s_b} \mid \bAstIfNd{s_b}{s_b} \mid \epsilon \quad\quad  \textit{BStmt} \ni s_b ::=\; \overrightarrow{b_b}
\end{align*}
\caption{
    The syntax of our formalised Viper subset (top, blue keywords) and corresponding Boogie subset (bottom, with subscript $b$, orange keywords) without top-level declarations.
    $\tau$ ($\tau_b$), $\mathit{bop}$, and $\mathit{uop}$ denote types, binary and unary operations, respectively.
}
\label{fig:viper_and_boogie_syntax}
\end{figure}

\subsection{The Viper and Boogie Languages}\label{subsec:viper_and_boogie_overview}
Viper programs in the subset considered here consist of a set of top-level declarations of fields (reference-field pairs are used to access the heap) and methods.
Boogie programs consist of a set of top-level declarations of global variables, constants, uninterpreted (polymorphic) functions, type constructors, axioms (which constrain the constants and functions), and procedures.
Both languages are imperative and separate \emph{statements} from \emph{expressions} (whose evaluation have no side-effects).
Viper additionally has separate \emph{assertions}.
The body of each Viper method and Boogie procedure is a statement.
Viper methods have pre- and post-conditions (assertions); method calls are verified modularly against these assertions.\footnote{Boogie supports pre-/post-conditions and procedure calls, but they are not used by the Viper-to-Boogie implementation.}
In Viper, scoped variables can be declared within statements; Boogie procedures declare all variables upfront.
Our supported Viper and Boogie statements, assertions, and expressions are shown in~\figref{fig:viper_and_boogie_syntax}.
Both languages have the same control flow elements and have some built-in types in common (\eg{} Booleans and integers).
Viper additionally provides a single \emph{reference} type, and supports reading from and writing to heap locations via a field access \vcode|e.f|, where \vcode{e} is a reference expression and \vcode{f} a field.

Our validation generates proofs that connect the abstract syntax tree (AST) of a Viper program (as represented by the Viper verifier) with the AST of the corresponding Boogie program (as represented by the Boogie verifier).\footnote{The Viper-to-Boogie implementation passes the Boogie program to the Boogie verifier via a text file. Targeting the Boogie AST as represented by the Boogie verifier in the proof avoids the need to trust the Boogie parser, and also generalises to verifier implementations that directly target the Boogie verifier's AST such as Dafny.}
Proof generation is complicated by the fact that the Viper and Boogie ASTs are structured differently. As shown in \figref{fig:viper_and_boogie_syntax}, the Viper AST uses a standard \emph{sequential composition} $s_1;s_2$, whereas
a Boogie statement is given by a list of \emph{statement blocks}.
Each statement block $\bAstBlock{\overrightarrow{c_b}}{\textit{if}_b}$ consists of a list of \emph{simple commands} (\ie{} no control flow), followed by either an if-statement or an empty statement ($\epsilon$).

As is typical for verifiers for higher-level languages, Viper's verification methodology employs a custom advanced program logic, in this case based on a flavour of separation logic (SL) called \emph{implicit dynamic frames} (IDF)~\cite{SmansJacobsPiessens12,ParkinsonSummers12} which reasons about the heap via \emph{permissions}.
Viper's assertions include the \emph{accessibility predicate} \vcode|acc(e.f,p)|, which represents a \emph{resource} (a logical notion which can be neither freely fabricated nor duplicated): the fractional (\vcode|p|) amount of \emph{permission to access heap location} \vcode{e.f}.\footnote{For readers familiar with separation logics, this is analogous to a fractional points-to assertion in a separation logic.}
Fractional permission amounts~\cite{Boyland03} range between 0 and 1; nonzero permission is required to \emph{read} heap locations and full (1) permission is required to \emph{write to} heap locations.
$\vsep{A}{B}$ expresses the \emph{separating conjunction} from SL, which specifies that the permissions in $A$ and $B$ must \emph{sum up to an amount currently held}.
One difference between IDF and SL is that IDF (and thus, Viper) supports general heap-dependent expressions such as \vcode{x.val = 5} or \code{x.f.f}, whose evaluation is \emph{partial} (only allowed with suitable permissions); this necessitates a notion of \emph{well-definedness} checks on expressions (see \secref{subsec:viper_semantics}).
Boogie does not provide built-in heap reasoning, and uses a much simpler program logic: its assertions are (total) formulas in first-order logic.

The presence of a heap in Viper also results in a very different state model. A Viper state consists of a variable store, a heap (mapping heap locations to current values) and a \emph{permission mask} (mapping heap locations to current permission amounts); a Boogie state is simply a variable store.

The main Viper features \emph{not} included in our subset are loops, more-complex resource assertions (predicates, magic wands, iterated separating conjunctions), heap-dependent functions, and domains.
Adding support for loops is straightforward: their semantics can be desugared via their invariant, in a pattern similar to method calls that we already support.
For other features more work would be required, but we are confident that these extensions would fit within our general methodology.

\subsection{Boogie Semantics}\label{subsec:boogie_semantics}
We extend our existing operational Boogie semantics formalised in Isabelle~\cite{ParthasarathyMuellerSummers21} to support the statements in~\figref{fig:viper_and_boogie_syntax}, and reuse many components including the state model and the semantics of simple commands.
The semantics of Boogie statements is expressed via program executions.
A finite program execution has one of three outcomes: (1)~it \emph{fails}, because an $\bassert{e}$ command is reached in a state that does not satisfy the Boolean expression $e$, (2)~it \emph{stops}, because an $\bassume{e}$ command is reached in a state that does not satisfy the Boolean expression $e$, or (3)~it \emph{succeeds}, because neither of the first two situations occur.
The three outcomes are represented formally via: (1)~a failure outcome \failureState{}, (2)~a \emph{magic} outcome \magicState{} for when the execution stops, and (3)~a \emph{normal} outcome \normalState{\stateBpl} in all other cases, where $\stateBpl$ is the resulting Boogie state, which is given by a mapping from variables to values.
Assignments and \bhavocNoArg{} commands always succeed; $\bhavoc{x}$ nondeterministically assigns a value of $x$'s declared type to $x$.

Formally, executions of Boogie statements are expressed via a small-step semantics.
The judgement $\redStmtBpl{\ctxtBpl}{\progPointBpl}{\normalState{\stateBpl}}{\progPointBpl'}{\stateExtBpl}$ expresses a finite execution \wrt{} \emph{Boogie context} $\ctxtBpl$ that takes 0 or more steps starting from the \emph{program point} $\progPointBpl$ and Boogie state $\stateBpl$, and ending in the program point $\progPointBpl'$ and outcome $\stateExtBpl$.
A Boogie context includes the interpretation of uninterpreted types and functions, and the types of declared variables.
A program point is given by a pair of the currently active statement block $b$ and the continuation representing the statement to be executed after $b$.
A continuation is either the empty continuation (\ie{} nothing to execute) or a sequential continuation (\ie{} a statement block followed by a continuation).
A continuation-based small-step semantics avoids the need for local search rules commonly required in a small-step semantics \cite{AppelB07}.

\subsection{Viper Semantics}\label{subsec:viper_semantics}
To our knowledge, there is no mechanised semantics for any fragment of the Viper language; we outline the main points of the one we have formalised here.
We give a big-step operational semantics to Viper statements via program executions again with three possible outcomes for finite executions: \emph{failure} \failureState{}, \emph{magic} \magicState{}, and \emph{normal outcomes} \normalState{\stateVpr} where $\stateVpr$ is a \emph{Viper state}.
A Viper state $\stateVpr$ comprises a local variable mapping \store{\stateVpr}, a heap \heap{\stateVpr} (a total mapping from heap locations to values), and a permission mask \mask{\stateVpr} (a total mapping from heap locations to permission amounts). The judgement $\redStmtVpr{\ctxtVpr}{s}{\stateVpr}{\stateExtVpr}$ holds if\gout{,} in the \emph{Viper context} $\ctxtVpr$ (fixing the declarations of methods, fields and local variables) the execution of statement $s$ in the state $\stateVpr$ terminates with outcome $\stateExtVpr$. 
Determining the outcome of a Viper execution is more involved than for Boogie as we will see below for the \vinhaleNoArg{} and \vexhaleNoArg{} operations.
Our semantics takes care that all Viper states are \emph{consistent}, \ie{} have \emph{consistent permission masks} (mapping each location to values between 0 and 1); executions that would produce inconsistent states in this sense are pruned by going to \magicState.


Formalising expression evaluation requires care for Viper, since, in a given state, not even all type-correct expressions are \emph{well-defined}: in our subset this can be either because of (1)~division by zero, or (2)~dereferencing a heap location for which no permission is held (subsuming null dereferences). 
In our semantics, evaluating an ill-defined expression causes execution to fail (in contrast to Boogie, where expression evaluation cannot fail).
Our judgement $\redExprVpr{e}{\stateVpr}{\normalVal{v}}$ expresses that expression $e$ evaluates to a value $v$ in state $\stateVpr$ (in particular, $e$ is well-defined in $\stateVpr$) and $\redExprVpr{e}{\stateVpr}{\failureVal{}}$ expresses that $e$ is ill-defined in $\stateVpr$.

Viper uses two main primitives to encode separation logic reasoning: (1)~$\vinhale{A}$ adds the permissions specified by assertion $A$ to the state, and \emph{stops} any execution where either a logical constraint in $A$ does not hold (these are \emph{assumed}) or the added permissions would yield an inconsistent state. (2)~$\vexhale{A}$ \emph{removes} the permissions specified by $A$, and \emph{fails} if either insufficient permissions are held or if a constraint in $A$ does not hold; for any heap locations to which \emph{all permission was removed}, an $\vexhaleNoArg{}$ also non-deterministically assigns arbitrary values.\footnote{For separation-logic-versed readers, the Hoare triples $\{R\}\;\vinhale{A}\;\{R*A\}$ and $\{R*A\}\;\vexhale{A}\;\{R\}$ reflect this behavior (assuming the expressions in $A$ and $R$ are well-defined).} 
This non-deterministic assignment reflects the fact that, while our Viper states employ total heaps (typical for IDF \cite{ParkinsonSummers12}), the values stored in heap locations without permission should be unconstrained.

$\vinhaleNoArg{}$ and $\vexhaleNoArg{}$ operations are typically used in Viper to encode external or more-complex operations \cite{MuellerSchwerhoffSummers16}. For instance, a Viper method call is expressed by exhaling the precondition and then inhaling the postcondition of the callee; the nondeterministic assignments made by the $\vexhaleNoArg{}$ model possible side effects of the call.
We present here some of the key rules for \vexhaleNoArg{}, which will be used later in this paper. Additional rules for $\vinhaleNoArg{}$ are presented in the appendix (\appreftr{app:inhale_semantics}{A}); the complete rules are included in our Isabelle formalisation.

\begin{figure}
\begin{align*}
%
%
&\Inf[\stmtExhSuccSemRuleName]{\redExhAux{\stateVpr}{A}{\stateVpr}{\normalState{\stateVpr'}}}.
{\exhNonDetSelect{\stateVpr}{\stateVpr'}{\stateVpr''}}
{\redStmtVpr{\ctxtVpr}{\vexhale{A}}{\stateVpr}{\normalState{\stateVpr''}}} \quad\quad
%
%
\Inf[\stmtExhFailSemRuleName]{\redExhAux{\stateVpr}{A}{\stateVpr}{\failureState{}}}
{\redStmtVpr{\ctxtVpr}{\vexhale{A}}{\stateVpr}{\failureState{}}}\\[3mm]
%
& \Inf[\exhAuxSepSemRuleName]{\redExhAux{\stateVpr^0}{A}{\stateVpr}{\normalState{\stateVpr'}}}.
{\redExhAux{\stateVpr^0}{B}{\stateVpr'}{\stateExtVpr}}.
{\redExhAux{\stateVpr^0}{A*B}{\stateVpr}{\stateExtVpr}}
\quad
%
%
\Inf[\exhAuxAccSemRuleName]{\redExprVpr{e}{\stateVpr^0}{\normalVal{r}}}
{\redExprVpr{e_p}{\stateVpr^0}{\normalVal{p}}}.
{\stateExtVpr = \textit{if} \; \exhAccSucc{r}{p}{\stateVpr}\;\textit{then}\;\normalState{\stateVpr^R} \;\textit{else}\; \failureState{}}
{\redExhAux{\stateVpr^0}{\vaccpred{e}{f}{e_p}}{\stateVpr}{\stateExtVpr}} \\[3mm]
%
%
&\exhNonDetSelect{\stateVpr}{\stateVpr'}{\stateVpr''} \eqdef{} 
\begin{array}{l}
  \store{\stateVpr''} = \store{\stateVpr'} \wedge{} \pi(\stateVpr'') = \pi(\stateVpr') \wedge{} \\ \forall l.\; (\pi(\stateVpr)(l) = 0 \vee \pi(\stateVpr')(l) > 0) \Rightarrow h(\stateVpr'')(l) = h(\stateVpr')(l)
\end{array} \\
& {\exhAccSucc{r}{p}{\stateVpr}} \eqdef{} p \geq 0 \wedge (r = \vnull{}\;?\; p = 0 : \mask{\stateVpr}(\vfieldacc{r}{f}) \geq p) \quad  \stateVpr^R \eqdef{} \removePerm{\stateVpr}{r}{f}{p}
\end{align*}
\caption{A subset of the rules for the formal semantics of exhale. $\removePerm{\stateVpr}{r}{f}{p}$ is the state $\stateVpr$ where permission $p$ is removed from $\vfieldacc{r}{f}$.}
\label{fig:semantics_exhale}
\end{figure}


An $\vexhale{A}$ must cause the loss of heap value information (via non-deterministic assignments) in general, but also needs to check that logical constraints \emph{were} true when the exhale started. Our semantics for 
$\vexhale{A}$ first removes the permissions and checks the constraints specified in $A$ \emph{without changing the heap yet} via an intermediate operation $\vexhaleAux{A}$; only then does it apply nondeterministic assignments. The inference rule~\stmtExhSuccSemRuleName{} in \figref{fig:semantics_exhale} formalises this behaviour for the case when $\vexhale{A}$ succeeds.
The big-step judgement $\redExhAux{\stateVpr}{A}{\stateVpr}{\normalState{\stateVpr'}}$ defines the successful execution of a $\vexhaleAux{A}$ operation from $\stateVpr$ to $\stateVpr'$.
\exhNonDetSelectNoArg{} specifies the nondeterministic assignment for all heap locations for which $\vexhaleAux{A}$ removed all permission.
The case when $\vexhaleAux{A}$ (and thus $\vexhale{A}$) fails, is captured by the rule \stmtExhFailSemRuleName{}.

Our semantics for $\vexhaleAux{A}$ decomposes the assertion $A$ from left to right:
That is, $\vexhaleAuxNoArg{}$ $\vsep{A}{B}$ first executes $\vexhaleAux{A}$ and then $\vexhaleAux{B}$ (rule~\exhAuxSepSemRuleName{} formalises the case when $\vexhaleAux{A}$ succeeds; if $\vexhaleAux{A}$ fails, then $\vexhaleAux{\vsep{A}{B}}$ also fails).
However, we need to also take care that the removal of permissions on-the-fly doesn't cause subexpressions to be considered ill-defined, e.g.~for the subexpression $\vfieldacc{x}{f}==1$  in $\vexhaleAux{\vsep{\vaccpred{x}{f}{1}}{\vfieldacc{x}{f} == 1}}$
which comes after the permission to $\vfieldacc{x}{f}$ is removed. 
Thus, our judgement carries both an \emph{\exhAuxExprEvalState} ($\stateVpr^0$ in~\exhAuxSepSemRuleName{}) in which expressions are evaluated and a \emph{\exhAuxReductionState} ($\stateVpr$ and $\stateVpr'$ in~\exhAuxSepSemRuleName{}) from which permissions are removed.
Rule~\exhAuxAccSemRuleName{} for $\vexhaleAux{\vaccpred{e}{f}{e_p}}$ models removing $e_p$ permission for heap location $\vfieldacc{e}{f}$. The operation succeeds (expressed by $\exhAccSucc{r}{p}{\stateVpr}$) iff (1)~the to-be-removed permission is nonnegative and, (2)~there is sufficient permission.
Rule~\exhAuxAccSemRuleName{} is applicable only if $e$ and $e_p$ are well-defined; there is a separate rule (not shown here) expressing that $\vexhaleAux{\vaccpred{e}{f}{e_p}}$ fails if $e$ or $e_p$ are ill-defined.

\subsection{Example Viper-to-Boogie Translation}
\label{subsec:vpr_to_bpl_example}

To give a flavour of a translation of a Viper statement into a Boogie statement, consider~\figref{fig:background_vpr_to_bpl_example}, which shows a simplified translation used by the existing Viper-to-Boogie implementation.
The Viper statement first adds permission to \code{x.f}, then updates \code{y.g}, and finally removes the added permission to \code{x.f} and checks that \code{y.g} is greater than \code{x.f}. This sequence of operations occurs, for instance, when verifying a method with the permission to \code{x.f} as precondition, the field update as method body, and the exhaled assertion as postcondition.

\begin{figure}
\small
\begin{tabular}{ll}
\begin{lstlisting}[language=silver]
inhale acc(x.f, q)
y.g := x.f+1
exhale acc(x.f, q) * y.g > x.f
\end{lstlisting}
& \translatesto \qquad
\begin{lstlisting}[language=boogie,numbers=left, firstnumber=1, stepnumber=1,mathescape]
tmp := q; assert tmp >= 0; $\label{line:background_inh_start}$
assume tmp > 0 ==> x != null;
M[x,f] += tmp;
assume GoodMask(M); $\label{line:background_inh_end}$
assert M[x,f] > 0; assert M[y,g] == 1; $\label{line:background_fieldassign_start}$
H[y,g] := H[x,f]+1;
assume GoodMask(M); $\label{line:background_fieldassign_end}$
WM := M; $\label{line:background_exh_start}$
tmp := q; assert tmp >= 0; $\label{line:background_exh_acc_nonfail_check_start}$
if(tmp != 0) {
  assert M[x,f] >= tmp;
} $\label{line:background_exh_acc_nonfail_check_end}$
M[x,f] -= tmp; $\label{line:background_exh_acc_end}$
assert WM[y,g] > 0; assert WM[x,f] > 0; $\label{line:background_exh_pure_start}$
assert H[y,g] > H[x,f]; $\label{line:background_exh_pure_end}$
havoc H'; assume idOnPositive(H,H',M); $\label{line:background_exh_non_det_select}$
H := H'; $\label{line:background_exh_heap_assignment}$
assume GoodMask(M); $\label{line:background_exh_end}$
\end{lstlisting}
\end{tabular}
\caption{A Viper statement (on the left) and the corresponding (simplified) Boogie statement (on the right) that is emitted by the current Viper-to-Boogie implementation.}
\label{fig:background_vpr_to_bpl_example}
\end{figure}

The corresponding Boogie program is significantly larger.
The \vinhaleNoArg{} is encoded on lines~\ref{line:background_inh_start}-\ref{line:background_inh_end}, the assignment is encoded on lines~\ref{line:background_fieldassign_start}-\ref{line:background_fieldassign_end}, and the \vexhaleNoArg{} is encoded on lines~\ref{line:background_exh_start}-\ref{line:background_exh_end}.
The Boogie program uses map-typed variables \bcode|H| and \bcode|M| to model the heap and permissions, respectively.\footnote{The notation \bcode|m[a]| is syntactic sugar here. We describe in~\secref{sec:evaluation} how maps are represented using the subset from~\figref{fig:viper_and_boogie_syntax}.}
The uninterpreted function \bcode|GoodMask| expresses when a permission mask is consistent; an axiom constrains the function correspondingly.
The permission mask of the expression evaluation state during the \vexhaleAuxNoArg{} operation is captured by the auxiliary variable \bcode{WM} (line~\ref{line:background_exh_start}). All locations in the assertion are checked to have positive permission \wrt{} \bcode{WM}.
The corresponding nondeterministic assignment of heap values is performed on lines~\ref{line:background_exh_non_det_select}-\ref{line:background_exh_heap_assignment}, where a heap \bcode{H'} is nondeterministically obtained via \bhavoc{\bcode{H'}} and then constrained to match the original heap \bcode{H} on all locations where there is positive permission (\wrt{} \bcode{M}) via the \bassumeNoArg{} statement; an axiom constrains the uninterpreted function \bcode{idOnPositive} correspondingly.
Note that this Boogie encoding overapproximates the nondeterministic assignment specified by
the Viper semantics: assigning new values to all locations without permission, rather than only
those newly without permission. 
Even this tiny snippet of code illustrates the explosion in concerns, complexity and the inobvious mapping between concepts in one language and the other, all of which must be taken care of in a formal validation approach.

\section{A Forward Simulation Methodology for Front-End Translations}
\label{sec:methodology}
A front-end translation is \emph{sound} iff the correctness of an input program is implied by the correctness of the correspondingly-translated IVL program. In our setting:
%
a Viper program (resp.\ a Boogie program) is \emph{correct} if each of its methods (resp.\ procedures) is correct.
At a high level (details in \secref{subsec:proof_strategy}), a method (resp.\ procedure) is correct if its body has no failing executions. 
Thus, proving soundness of the Viper-to-Boogie translation boils down to proving that \emph{if} the Viper program has a failing execution, then the translated Boogie program has one also.

We generate such proofs \emph{automatically} via a novel general methodology for proving \emph{forward simulations}~\cite{LynchV95} between source and IVL target statements. We observed early on that generating such proofs directly based on knowledge of the entire translation would require handling the entire semantic gap between the source and target languages monolithically in one result, which would be both infeasible to automate effectively and highly-brittle to any changes in the translation.

Instead, our methodology employs a combination of key strategies that work together to achieve reliable and robust automation of our formal simulation results: (1) syntactic and semantic \emph{decompositions} into smaller and more-focused simulation sub-results that are easier to automate, (2) \emph{generic simulation judgements} which can be instantiated to obtain the diverse simulation notions we require, (3) \emph{generic composition lemmas} which factor out common idioms arising in diverse facets of the translation, and (4) \emph{contextual hypotheses} which can be injected into simulation proofs to handle non-locality of certain translation checks. We present these key ingredients of our methodology in this section. We illustrate them for Viper and Boogie, but they can be naturally ported to other front-end translations if one provides a formal semantics for the input language and IVL, because they are designed to abstract over states, relations and statements employed in a translation.
%

\subsection{Focusing Forward Simulation Proofs by Decomposition}
\label{subsec:forward_simulation}
Intuitively, a forward simulation between a Viper and a Boogie statement shows that for any execution of the Viper statement, there exists a corresponding execution of the Boogie statement that \emph{simulates} it. By defining the simulation such that 
a \emph{failing} Viper execution is simulated only by \emph{failing} Boogie executions, a forward simulation implies our desired result in particular.

To tackle the complexity of automatically (and reliably) generating simulation proofs in general for the Viper-to-Boogie translation, we employ a variety of strategies for aggressively decomposing the desired simulation result into smaller and simpler sub-goals that are themselves still simulation results. These decompositions are sometimes intuitive based on the syntax: for example, in the case of decomposing simulation of a Viper sequential composition into simulations for its constituent statements. However, we go \emph{further than the syntax}, decomposing across different \emph{semantic concerns} for the \emph{same} Viper statement, into what we call Viper \emph{effects}.

For example, we discussed in \secref{subsec:viper_semantics} that the semantics of $\vexhaleNoArg{}$ consists of two effects, \vexhaleAuxNoArg{} and a nondeterministic assignment. The simulation proofs for each of these Viper effects are made separately, and then composed for a simulation proof for the primitive statement as a whole; this would in turn be composed with simulation proofs for other sequentially-composed statements, and so on. Note in particular, that simulation proofs may need to relate only a \emph{part of} the semantics of a Viper statement to some appropriate Boogie code, a technicality which requires special care when tracking the \emph{relations} between corresponding states in the two programs.

Via our decompositions, each resulting simulation proof focuses on a different specific semantic concern with respect to the translation in question; these proofs can be made simple enough to discharge automatically, optionally with tailored tactics. However without care, our decomposition approach could lead easily to an explosion of ad hoc simulation judgements with disparate forms and parameters. Instead, our simulation methodology defines a \emph{single, generic} simulation judgement which can be instantiated appropriately to define each particular simulation judgement required. We design our generic judgements to support instantiations which reflect not only the semantics of the particular effect in isolation, but to optionally include additional contextual information to be propagated to specialise and aid the simulation proof itself.


\begin{figure*}
\begin{align*}
%
%
&\genericSim{\ctxtBpl}{R_{\mathit{in}}}{R_{\mathit{out}}}{\mathit{Succ}}{\mathit{Fail}}{\progPointBpl_{\mathit{in}}}{\progPointBpl_{\mathit{out}}} \eqdef{} \forall \tau, \stateBpl.\; R_{\mathit{in}}(\tau, \stateBpl) \Longrightarrow \\
&\quad \left(\forall \tau'.\; \mathit{Succ}(\tau, \tau') \Longrightarrow \exists \stateBpl'.\; \redStmtBpl{\ctxtBpl}{\progPointBpl_{\mathit{in}}}{\normalState{\stateBpl}}{\progPointBpl_{\mathit{out}}}{\normalState{\stateBpl'}} \wedge R_{\mathit{out}}(\tau',\stateBpl') \right) \wedge{} \tag{Success case} \\
&\quad \left(\mathit{Fail}(\tau) \Longrightarrow \exists \progPointBpl'.\; \redStmtBpl{\ctxtBpl}{\progPointBpl_{\mathit{in}}}{\stateBpl}{\progPointBpl'}{\failureState{}} \right) \tag{Failure case} \\[0.5em]
%
%
&\stmtSim{\ctxtVpr}{\ctxtBpl}{R}{R'}{s}{\progPointBpl}{\progPointBpl'} \eqdef \\
& \quad \genericSim{\ctxtBpl}{R}{R'}{\lambda \stateVpr\;\stateVpr'.\;\redStmtVpr{\ctxtVpr}{s}{\stateVpr}{\normalState{\stateVpr'}}}{\lambda \stateVpr.\;\redStmtVpr{\ctxtVpr}{s}{\stateVpr}{\failureState{}}}{\progPointBpl}{\progPointBpl'}\\
%
%
&\exprsWfSim{\ctxtBpl}{R}{R'}{es}{\progPointBpl}{\progPointBpl'} \eqdef{} \genericSimNoArg_{\ctxtBpl}
\left(
\begin{array}{l}
R,R', (\lambda \stateVpr\; \stateVpr'.\; \stateVpr = \stateVpr' \wedge \exists vs.\; \redExprsVpr{es}{\stateVpr}{\normalVal{vs}}), \\
(\lambda \stateVpr.\;\redExprsVpr{es}{\stateVpr}{\failureVal{}}),\progPointBpl,\progPointBpl'
\end{array}
\right)\\
%
%
%
&\exhAuxSim{\ctxtBpl}{R}{R'}{A}{\progPointBpl}{\progPointBpl'} \eqdef{} \genericSimNoArg_{\ctxtBpl}
\left(
\begin{array}{l}
R,R', (\lambda (\stateVpr^0,\stateVpr)\;(\stateVpr^1,\stateVpr').\;\stateVpr^0 = \stateVpr^1 \wedge \redExhAux{\stateVpr^0}{A}{\stateVpr}{\normalState{\stateVpr'}}), \\
(\lambda (\stateVpr^0,\stateVpr).\;\redExhAux{\stateVpr^0}{A}{\stateVpr}{\failureState{}}),\progPointBpl,\progPointBpl'
\end{array}
\right)
\end{align*}
\caption{The definition of the generic forward simulation judgement and three common instantiations. The judgement $\redExprsVpr{es}{\stateVpr}{r}$ lifts the evaluation of an expression (see~\secref{subsec:viper_semantics}) to a list of expressions $es$.}
\label{fig:forward_simulation}
\end{figure*}


\subsection{One Simulation Judgement to Rule Them All}
\label{subsec:generic_forward_simulation}

Our generic forward simulation judgement \genericSimNoArg{} is defined in~\figref{fig:forward_simulation}.
All concrete forward simulations (\eg{} for statements, well-definedness checks, etc.) are instantiations of this judgement. As well as aiding understanding, this approach enables both tactics which manipulate this generic judgement directly, and \emph{generic composition proof rules} which embody recurring proof idioms in a way which is again parametric with the specific simulations in question (\secref{subsec:inst_indep_rules}). 

\genericSimNoArg{} is defined in terms of multiple parameters: (1) the Boogie context $\ctxtBpl$, (2) an \emph{input relation} $R_{\mathit{in}}$ and an \emph{output relation} $R_{\mathit{out}}$ on Viper and Boogie states, (3) a \emph{success predicate} $\mathit{Succ}$ characterising the set of input and output Viper state pairs $(\tau,\tau')$ for which there is a successful Viper execution from $\tau$ to $\tau'$, (4) a \emph{failure predicate} $\mathit{Fail}$ characterising the set of input Viper states that result in a failing execution, (5) input and output Boogie program points $\progPointBpl_{\mathit{in}}$ and $\progPointBpl_{\mathit{out}}$ where the Boogie executions are expected to start and end, respectively.
The success and failure predicate together abstractly describe the set of Viper executions that must be shown to be simulated. 

$\genericSim{\ctxtBpl}{R_{\mathit{in}}}{R_{\mathit{out}}}{\mathit{Succ}}{\mathit{Fail}}{\progPointBpl_{\mathit{in}}}{\progPointBpl_{\mathit{out}}}$ holds iff for any Viper and Boogie input states related by $R_{\mathit{in}}$, the following two conditions hold: (1) for any successful Viper execution from the input Viper state to an output Viper state $\tau'$, there must be a Boogie execution from program point $\progPointBpl_{\mathit{in}}$ and the input Boogie state to program point $\progPointBpl_{\mathit{out}}$ and some output Boogie state that is related to $\tau'$ by $R_{\mathit{out}}$, and (2) if the Viper execution fails in the input state, then there must be a failing Boogie execution from  $\progPointBpl_{\mathit{in}}$ and the input Boogie state (the reached Boogie program point need not be $\progPointBpl_{\mathit{out}}$).
The second condition is the end goal that we need to show soundness of the Viper-to-Boogie translation.
The first condition is needed in order to derive \genericSimNoArg{} compositionally; it guarantees, for example, that not all Boogie executions for a successful Viper execution produce a magic outcome.

Three important instantiations of \genericSimNoArg{} that we use are shown at the bottom of~\figref{fig:forward_simulation}.
\stmtSimNoArg{} is the forward simulation for Viper statements, where the success and failure predicates are instantiated to be a successful and a failing Viper statement reduction, respectively. 
Thus, the resulting failure case in \genericSimNoArg{} directly gives us the key property to show the soundness of a Viper-to-Boogie translation.
\exprWfSimNoArg{} is the forward simulation for the well-definedness check of a list of Viper expressions.
Here, the instantiation of the success predicate explicitly expresses that the Viper state does not change during the evaluation of expressions.
\exhAuxSimNoArg{} is the forward simulation for \vexhaleAuxNoArg{}.
Here, the instantiation makes use of the fact that the generic simulation judgement \genericSimNoArg{} is in fact also (implicitly, here) parametric with the notions of states employed: the ``Viper state'' is in fact instantiated to be a pair of standard Viper states in this case, where the first Viper state represents the \exhAuxExprEvalState{} and the second Viper state represents the \exhAuxReductionState{} (see~\secref{subsec:viper_semantics} for this distinction).
The success predicate expresses that the \exhAuxExprEvalState{} does not change during a \vexhaleAuxNoArg{} operation.
These three common instantiations are all expressed directly via the Viper reduction judgements introduced in~\secref{subsec:viper_semantics}.
Like the generic simulation judgement, the three instantiations \emph{are themselves generic}, abstracting away \emph{how} the Viper and Boogie states are related by taking the input and output state relations as parameters.
As we will show in~\secref{subsec:modularity_and_abstraction}, we also use instantiations that do not just use Viper reduction judgements (\eg{} to express the non-deterministic assignment of heap values in~\vexhaleAuxNoArg{}). 

\subsection{Instantiation-Independent Rules}
\label{subsec:inst_indep_rules}
\begin{figure}
\begin{align*}
&\Inf[\compRuleName]{\genericSim{\ctxtBpl}{R}{R'}{S_1}{F_1}{\progPointBpl}{\progPointBpl'}}.
{\genericSim{\ctxtBpl}{R'}{R''}{S_2}{F_2}{\progPointBpl'}{\progPointBpl''}}.
{\forall \tau,\tau''.\; S(\tau,\tau'') \Rightarrow \exists \tau'.\; S_1(\tau,\tau') \wedge S_2(\tau',\tau'') }.
{\forall \tau.\; F(\tau) \Rightarrow F_1(\tau) \vee \exists \tau'.\; S_1(\tau,\tau') \wedge F_2(\tau') }.
{\genericSim{\ctxtBpl}{R}{R''}{S}{F}{\progPointBpl}{\progPointBpl''}}
\;
\Inf[\propRuleName]{\onlyBoogieSim{\ctxtBpl}{R}{R_1}{\progPointBpl}{\progPointBpl_1}}.
{\genericSim{\ctxtBpl}{R_1}{R_2}{S}{F}{\progPointBpl_1}{\progPointBpl_2}}.
{\onlyBoogieSim{\ctxtBpl}{R_2}{R'}{\progPointBpl_2}{\progPointBpl'}}.
{\genericSim{\ctxtBpl}{R}{R'}{S}{F}{\progPointBpl}{\progPointBpl'}} \\[3mm]
&\Inf[\stmtSeqRuleName]{\stmtSim{\ctxtVpr}{\ctxtBpl}{R}{R'}{s_1}{\progPointBpl}{\progPointBpl'}}.
{\stmtSim{\ctxtVpr}{\ctxtBpl}{R'}{R''}{s_2}{\progPointBpl'}{\progPointBpl''}}.
{\stmtSim{\ctxtVpr}{\ctxtBpl}{R}{R''}{(s_1;s_2)}{\progPointBpl}{\progPointBpl''}}
\qquad
\textit{where } 
\begin{array}{l}
\onlyBoogieSim{\ctxtBpl}{R}{R'}{\progPointBpl}{\progPointBpl'} \eqdef{} \\[3mm]
\genericSim{\ctxtBpl}{R}{R'}{\lambda \tau\;\tau'.\; \tau = \tau'}{\lambda \_.\; \bot}{\progPointBpl}{\progPointBpl'}  
\end{array}
\end{align*}
\caption{The instantiation-independent rules \compRuleName{} and \propRuleName{} and the concrete rule for the simulation of $s_1;s_2$.}
\label{fig:generic_forward_simulation_rules}
\end{figure}

Many simulation idioms arise repeatedly in a complex translation. Notions of sequential composition, conditional evaluation, stuttering steps are all good examples, which require a certain stylised formal justification to reason about. Our generic simulation judgement allows us to identify and formalise these idioms once and for all, providing, for example, generic composition lemmas that can be proved once and instantiated for different purposes. In this subsection, we present these idioms as inference rules, but in our formalisation they are expressed and proved as regular lemmas.

For example, we prove a single general composition rule from which we derive 
concrete rules to combine (1)~simulations of $s_1$ and $s_2$ to a simulation of $s_1;s_2$, (2)~simulations of $\vexhaleAux{A_1}$ and $\vexhaleAux{A_2}$ to $\vexhaleAux{\vsep{A_1}{A_2}}$, (3)~simulations of $\vinhale{A_1}$ and $\vinhale{A_2}$ to $\vinhale{\vsep{A_1}{A_2}}$. 
The general composition rule~\compRuleName{} in~\figref{fig:generic_forward_simulation_rules} captures the composition of two, possibly different, instantiations of \genericSimNoArg{}, where the output relation and Boogie program point of the first instantiation match the input relation and program point of the second one.
The two final premises constrain the resulting success and failure predicates.
In particular, the composed Viper execution should fail only if either the first instantiation fails or if the second instantiation fails in a state successfully reached by the first one.
The rule~\stmtSeqRuleName{} in~\figref{fig:generic_forward_simulation_rules} shows the concrete composition rule for $s_1;s_2$, which is derived from~\compRuleName{}. 
Note that \stmtSeqRuleName{} does not impose any constraints on the Boogie program points, which is crucial to handle Viper's and Boogie's disparate ASTs (see \secref{subsec:viper_and_boogie_overview}).
We will discuss in \secref{subsec:proof_automation} how we deal with the AST mismatch when automating proofs.

As a second example, the notion of simulation \emph{stuttering steps} also arises in many ways, whenever some auxiliary Boogie code is generated that does not fully correspond to a step in the Viper source. This includes initialisations of auxiliary variables, or Boogie \bassumeNoArg{} statements for properties from the current simulation state relation. This idiom is captured by the \emph{Boogie propagation rule}~\propRuleName{} in~\figref{fig:generic_forward_simulation_rules}, in which \onlyBoogieSimNoArg{} expresses simulations in which the Viper state remains unchanged, and thus only the Boogie state may change (causing adjustment to the state relations).

\subsection{Examples: Generic Decomposition in Action}
\label{subsec:modularity_and_abstraction}

As outlined above, the general strategy for our simulation methodology is to decompose our simulation goals as far as possible, while leaving as many parameters generic as we can to enable maximal reuse of our results and composition lemmas. While decomposition handles the semantic gap, our use of generic parameterisation provides the abstraction \gout{necessary}to address the diverse translations used in practical translational verifiers.
In the following, we showcase our methodology on one rule, but the same ideas apply to all our formal rules (see a second example in~\appreftr{app:simulation_rule_example}{B}).

\begin{figure}
\begin{align*}
&\Inf[\stmtSimExhRuleName]
{\exhAuxSim{\ctxtBpl}{[\lambda (\stateVpr^0,\stateVpr)\;\stateBpl.\; \stateVpr^0 = \stateVpr \wedge R(\stateVpr, \stateBpl)]}
{R'}{A}{\progPointBpl}{\progPointBpl'}}{\; (\text{sim. of } \vexhaleAux{A})}.
{\genericSim{\ctxtBpl}{R'}{[\lambda (\_,\stateVpr)\;\stateBpl.\; R''(\stateVpr, \stateBpl)]}{\mathit{Succ}_2}{\lambda \_.\; \bot}{\progPointBpl'}{\progPointBpl''}}{\; \text{(non-det. selection)}}
{\stmtSim{\ctxtVpr}{\ctxtBpl}{R}{R''}{\vexhale{A}}{\progPointBpl}{\progPointBpl''}} \\[0.8em]
&\mathit{Succ}_2 \eqdef{} \lambda (\stateVpr^0,\stateVpr)\;(\_,\stateVpr').\;  \exhNonDetSelect{\stateVpr^0}{\stateVpr}{\stateVpr'} \wedge{} \redExhAux{\stateVpr^0}{A}{\stateVpr^0}{\stateVpr}
\end{align*}
\caption{Rule for the simulation of $\vexhale{A}$. The definition of \exhNonDetSelectNoArg{} is given in~\figref{fig:semantics_exhale}.}
\label{fig:stmt_exh_simulation}
\end{figure}

%

Consider the rule~\stmtSimExhRuleName{} for the simulation of~\vexhale{$A$} in \figref{fig:stmt_exh_simulation}. The first premise is expressed as a simulation of the first effect, \vexhaleAuxNoArg{$A$}, which we can express via the \exhAuxSimNoArg{} instantiation (see~\figref{fig:forward_simulation}).
The second premise models nondeterministic assignment, which is captured by the first conjunct \exhNonDetSelectNoArg{} of the corresponding success predicate and by the failure predicate, which reflects that the nondeterministic assignment cannot fail.

By modularly abstracting over the details of these premises, and the precise definitions of the states and state relations (\eg{} the intermediate relation $R'$ in this rule), we obtain robustness to diverse translations: our rules do not constrain \emph{which exact Boogie statements} correspond to a Viper effect.
For example, the Viper-to-Boogie implementation establishes the nondeterministic heap assignment premise in~\stmtSimExhRuleName{} in two different ways depending on whether the assertion contains an accessibility predicate $\vaccpred{e}{f}{p}$ or not; if not, then the implementation does not emit any Boogie code for the nondeterministic assignment, which is sound, since no permission is removed.

%

Note that this genericity does not prevent the rule from exploiting contextual information.
For example, the input state relation of the first premise specifies that at the beginning of the $\vexhaleAux{A}$ effect the \exhAuxExprEvalState{} and the \exhAuxReductionState{} are the same.
This property does not hold in general for executions of $\vexhaleAuxNoArg{}$ (\eg{} it might not hold when executing the second conjunct of a separating conjunction), but it does hold here, at the beginning of an $\vexhaleNoArg{}$.
The second premise's success predicate includes the fact that the current Viper state was reached via \vexhaleAux{$A$}.
This allows us, for example, to conclude that the nondeterministic assignment has no effect if \vexhaleAux{$A$} removes no permissions, which is required to justify the case when the implementation does not emit Boogie code for the nondeterministic assignment.

\subsection{Injecting Non-Local Hypotheses into Simulation Proofs}
\label{subsec:properties_in_state_relation}

Our rules are designed to be parametric in the state relation between the Viper and Boogie state and permit adjusting this state relation at different points in the simulation proof (\eg{} via the Boogie propagation rule~\propRuleName{} in~\figref{fig:generic_forward_simulation_rules}). In principle, this allows the injection of arbitrary non-locally-justified hypotheses into all of our simulation judgements. However, automating the \emph{usage} of general logical assumptions embedded into our state relations can become a challenge in itself.


For example, the Viper-to-Boogie implementation omits the well-definedness checks of expressions in the translation of $\vexhaleAux{A}$ and $\vinhale{A}$ in certain cases (as we will discuss in \secref{subsec:vpr_to_bpl_method_calls}).
This is justified, because $A$ is checked to be \emph{well-formed} non-locally in those cases, but to \emph{use} this additional hypothesis requires propagating and adjusting it through the cases of the definition of $\vexhaleAux{A}$ and $\vinhale{A}$.


As a final ingredient of our methodology, to avoid these recurring adaptations and proof steps, we allow \emph{specialised} instantiations of the generic forward simulation judgement \genericSimNoArg{} that encapsulate these extra hypotheses as 
\emph{additional} premises. Thus, applications of the rule can work with a fixed state relation and replace recurring proof steps by the justification of an additional premise.

\begin{figure*}
\begin{align*}
%
&\Inf[\exhAuxInvSepRuleName]{\exhAuxInvSim{\ctxtBpl}{Q}{R}{R'}{A_1}{\progPointBpl}{\progPointBpl'}}
{\exhAuxInvSim{\ctxtBpl}{Q}{R'}{R''}{A_2}{\progPointBpl'}{\progPointBpl''}}.
{ \forall \stateVpr^0, \stateVpr.\; Q(A_1*A_2, (\stateVpr^0, \stateVpr)) \Rightarrow 
\left(
\begin{array}{l}
Q(A_1, (\stateVpr^0, \stateVpr)) \wedge{} \\
\forall \stateVpr'.\; \redExhAux{\stateVpr^0}{A_1}{\stateVpr}{\normalState{\stateVpr'}} \Rightarrow Q(A_2, (\stateVpr^0, \stateVpr'))
\end{array}
\right)
}
{\exhAuxInvSim{\ctxtVpr}{Q}{R}{R''}{(A_1*A_2)}{\progPointBpl}{\progPointBpl''}}\\[0.5em]
%
& \exhAuxInvSim{\ctxtBpl}{Q}{R}{R'}{A}{\progPointBpl}{\progPointBpl'} \eqdef{}
\exhAuxSim{\ctxtBpl}{(\lambda \tau, \stateBpl.\;R(\tau,\stateBpl) \wedge{} Q(A, \tau))}{R'}{A}{\progPointBpl}{\progPointBpl'} 
\end{align*}
\caption{The instantiation for simulating \vexhaleAux{$A$} with  assertion predicate $Q$ (bottom) and the corresponding rule for the separating conjunction (top).}
\label{fig:assertion_invariant}
\end{figure*}

For example, \figref{fig:assertion_invariant} shows (at the bottom) an instantiation of \genericSimNoArg{} that expresses the simulation of \vexhaleAux{$A$}, parameterised with a predicate $Q$ on assertions. 
Its definition in terms of \exhAuxSimNoArg{} requires $Q(A,\tau)$ to hold as part of the input state relation.
The specialised rule \exhAuxInvSepRuleName{} (top of \figref{fig:assertion_invariant}) for $\vexhaleAux{\vsep{A_1}{A_2}}$ 
decomposes the simulation into simulations for $A_1$ and $A_2$.\footnote{\exhAuxInvSepRuleName{} can be derived from the instantiation-independent composition rule (\figref{fig:generic_forward_simulation_rules}) and consequence rule~\conseqRuleName{} (\appreftr{app:conseq_rule}{C}).} Both sub-simulations use the \emph{same} predicate $Q$, such that applications of the rule do \emph{not} need to adjust the state relations explicitly to reflect that, for example, $Q$ holds for $A_1$ and $A_2$ in the respective states. This property is ensured by the third premise. In practice, for a specific $Q$, we prove the third premise once and for all for all assertions $A_1$ and $A_2$, which avoids the recurring proof steps that would be necessary without the specialised rule. Note that the same parameter can be instantiated in many ways to capture different non-local hypotheses for different applications of the same rule.

\bigskip
In summary, our methodology solves all three challenges outlined in the introduction. The \emph{large semantic gap} between the input language and the IVL is handled by decomposing the statements of the input language into smaller effects and defining for each of them instantiations of a generic forward simulation relation. The parameterisation of this relation allows us, in particular, to capture information about the context in which the effects are executed. This parameterisation also supports \emph{diverse translations} by abstracting from the details of the translation. Finally, \emph{non-locality} is handled by capturing properties checked elsewhere in the state relations, and by devising specialised rules that simplify the proof generation.
All of these ideas are needed to validate the existing Viper-to-Boogie translation, but apply equally to other front-end translations.

\section{Putting The Methodology to Work}\label{sec:validation_impl}
This section presents ideas for applying the methodology from~\secref{sec:methodology} to concrete front-end translations.
In particular, the section presents our instantiation of the state relation (\secref{subsec:state_relation_impl}), a concrete instance of non-local reasoning (\secref{subsec:vpr_to_bpl_method_calls}), and how our proof automation works (\secref{subsec:proof_automation}).
Finally, the section discusses the background theory for Boogie (\secref{subsec:background_theory}), which includes polymorphic maps, and shows how to use forward simulation proofs to generate the final theorem (\secref{subsec:proof_strategy}).

\subsection{State Relation}\label{subsec:state_relation_impl}
Our rules for deriving forward simulation judgements (\secref{sec:methodology}) allow us to adjust state relations as needed during a simulation proof. We use this flexibility in many ways, \eg{} when (1) a scoped Viper variable is introduced, (2) a new auxiliary Boogie variable is introduced, (3) the Boogie variables tracking the Viper state are changed.
To facilitate proof automation for handling such adjustments, we build in a stylised form for expressing state relations for this translation via two parameters: 
a partial \emph{\auxVarMap} from auxiliary Boogie variables to \emph{logical conditions they each satisfy}, and a \emph{translation record} specifying how key Viper components are represented in the Boogie state; the scenarios above are all handled by adjusting one of these two parameters.
Translation records comprise: (1) a mapping \varTr{\trRecord} from Viper variables to their Boogie counterparts, (2) the Boogie variables representing the Viper heap \heapTr{\trRecord} and permission mask \maskTr{\trRecord} (and whenever we use a separate expression evaluation state, the corresponding variables representing the heap \heapDefTr{\trRecord} and mask \maskDefTr{\trRecord})
and (3) a mapping \fieldTr{\trRecord} from Viper fields to corresponding Boogie constants.

The following definition shows a simplified excerpt of our state relation instantiation \stateRelInstName{} for translation record \trRecord{} and \auxVarMap{} \auxVars{}, where $\stateVpr$ and $\stateBpl$ are the Viper and Boogie states, and $\stateVpr^0$ is a distinguished Viper expression evaluation state (if there is none, then $\stateVpr = \stateVpr^0$):
\begin{align*}
&\stateRelInst{\ctxtBpl}{\trRecord}{\auxVars}{\stateVpr^0}{\stateVpr}{\stateBpl} \eqdef{} \consistent{\stateVpr^0} \wedge{} \consistent{\stateVpr} \wedge{} \\
&\fieldRel{\ctxtBpl}{\fieldTr{\trRecord}}{\stateBpl}  \wedge{} (\forall x,P.\; \auxVars(x) = P \Rightarrow P(\stateBpl(x))) \wedge{} \\
&\storeRel{\ctxtBpl}{\varTr{\trRecord}}{\stateVpr}{\stateBpl} \wedge{}
 \heapMaskRel{\ctxtBpl}{\heapTr{\trRecord}}{\maskTr{\trRecord}}{\stateVpr}{\stateBpl} \wedge{} 
\heapMaskRel{\ctxtBpl}{\heapDefTr{\trRecord}}{\maskDefTr{\trRecord}}{\stateVpr^0}{\stateBpl} \wedge{} ...
\end{align*}
The first line ensures that the Viper states are consistent. 
The second line ensures that the Viper fields are represented in the Boogie state (\fieldRelNoArg{}) and 
 that for each 
 $(x,P)$ in the \auxVarMap{}, $P$ holds for the value 
 of $x$. 
The third line ensures that the Boogie state correctly captures the Viper state: both in terms of its variable store (\storeRelNoArg{}) and heap and permission mask (\heapMaskRelNoArg{}).
%

\subsection{Non-Locality}\label{subsec:vpr_to_bpl_method_calls}
For most occurrences of \vexhaleAux{$A$}, the Viper-to-Boogie implementation generates well-\\definedness checks in the Boogie program corresponding to expressions evaluated in $A$. However, specifically when executing the \vexhaleNoArg{} of a method call's precondition, the translation omits these well-definedness checks for the corresponding \vexhaleAuxNoArg{} operation. This is justified by a \emph{non-local check}:  
the Boogie code for the \emph{callee}'s translation checks that the callee's specification is \emph{well-formed}, which implies that expressions evaluated in the precondition will always be well-defined.\footnote{There is an analogous non-local check for $m$'s postcondition that we do not discuss here for simplicity of presentation.}

Given Viper's semantics, our standard simulation proof for \vexhaleAux{$A$} would fail if we did not reflect the consequences of this non-local guarantee \emph{in a way that is used automatically during the proof}.
We instantiate the general strategy outlined in~\secref{subsec:properties_in_state_relation} for this purpose, which allows us to choose a predicate $Q_\mathit{pre}$ on assertions that will be applied throughout the simulation proof for \vexhaleAux{$A$}. Our strategy requires us to find $Q_\mathit{pre}$ such that (a) it is implied by the non-local check elsewhere, and (b) it can be propagated identically to sub-assertions of $A$ during the proof (\eg{} satisfying the third premise of~\exhAuxInvSepRuleName{} in~\figref{fig:assertion_invariant}, and similarly for other connectives).

In this case, we instantiate the predicate in our strategy with the following definition: 
$$
Q_\mathit{pre}(A, \stateVpr^0, \stateVpr) \eqdef{} \consistent{\stateVpr^0} \wedge{} \exists \stateVpr^{\mathit{i}}.\; \stateVpr \oplus \stateVpr^{\mathit{i}} \leqVpr \stateVpr^0 \wedge{} \neg{\redInh{A}{\stateVpr^{\mathit{i}}}{\failureState{}}} 
$$
Here, \redInh{A}{\stateVpr}{\stateExtVpr} holds iff \redStmtVpr{\ctxtVpr}{\vinhale{A}}{\stateVpr}{\stateExtVpr},
and $\oplus$ and $\leqVpr$ (and later, $\ominus$) have standard pointwise meanings on permission masks, leaving heaps and stores identical. This predicate expresses that possibly after restoring some permissions (in $\stateVpr^{\mathit{i}}$) that we \emph{had} at the start of the \vexhaleNoArg{}, at least an \vinhaleNoArg{} of $A$ would not fail (\ie{} expressions evaluated within $A$ will be well-defined). 
The non-local check of the method precondition, which effectively checks that an \vinhaleNoArg{} would not fail starting from an \emph{empty} state (\ie{} no permissions), implies the predicate for an empty $\stateVpr^{\mathit{i}}$.
Showing formally that $Q_\mathit{pre}$ can be propagated over connectives occurring in $A$ requires in particular a technical lemma stating a partial \emph{inversion} property between \vexhaleAuxNoArg{} and \vinhaleNoArg{}:
\begin{lemma}
Let $A$ be an assertion and $\stateVpr^0$, $\stateVpr'$, $\stateVpr^i$, $\stateVpr^s$ be Viper states, where $\stateVpr^s = \stateVpr^i \oplus (\stateVpr \ominus \stateVpr')$ and $\stateVpr^s$ is consistent. If $\redExhAux{\stateVpr^0}{A}{\stateVpr}{\normalState{\stateVpr'}}$ and $\neg{\redInh{A}{\stateVpr^{\mathit{i}}}{\failureState{}}}$ holds, then $\redInh{A}{\stateVpr^{\mathit{i}}}{\normalState{\stateVpr^s}}$.
\label{lemma:exhale_inhale_relation}
\end{lemma}
We prove this result by induction on the reduction of $\vexhaleAuxNoArg{}$.
The lemma essentially states that the permissions that get removed by $\vexhaleAux{A}$ (expressed by $\stateVpr \ominus \stateVpr'$) are exactly those that will be added by a corresponding (non-failing) $\vinhale{A}$ operation.

\subsection{Proof Automation}\label{subsec:proof_automation}
\begin{figure}
\begin{align*}
&\text{Proof Tree $T$}: \quad {
    \Inf[\exhAuxNoInvSepRuleName]{
      \Inf{\fbox{
        \text{Proof $\mathcal{P}_2$ (hint 3)}
      }}
      {
        \exhAuxSim{\ctxtBpl}{R_2}{R_2}{A_1}{\progPointBpl_1}{\progPointBpl_2} 
      }
    }{
      \Inf{\fbox{
          \text{Proof $\mathcal{P}_3$ (no hint)}
      }}{
        \exhAuxSim{\ctxtBpl}{R_2}{R_2}{A_2}{\progPointBpl_2}{\progPointBpl_3} 
      }
    }
    {
      \exhAuxSim{\ctxtBpl}{R_2}{R_2}{\vsep{A_1}{A_2}}{\progPointBpl_1}{\progPointBpl_3} 
    }
} \\
&\Inf[\underbrace{\stmtSimExhRuleName}_{\text{hint 1}}]
{
  \Inf[\exhAuxPropRuleName]{
    \Inf{
      \fbox{
        \text{Proof $\mathcal{P}_1$ (hint 2)}
      }
    }{
      \onlyBoogieSim{\ctxtBpl}{R_1}{R_2}{\progPointBpl}{\progPointBpl_1} 
    }
  }{(\text{Proof Tree $T$})}
  { \exhAuxSim{\ctxtBpl}{R_1}{R_2}{\vsep{A_1}{A_2}}{\progPointBpl}{\progPointBpl_3} }
}
{
  \Inf{
  \fbox{ \text{Proof $\mathcal{P}_4$ (hints 4 and 5)} }
  }{
  \genericSim{\ctxtBpl}{R_2}{R_3}{\mathit{Succ}_2}{\lambda \_.\; \bot}{\progPointBpl_3}{\progPointBpl'}
  }
}
{
\stmtSim{\ctxtVpr}{\ctxtBpl}{R}{R}{\vexhale{\vsep{A_1}{A_2}}}{\progPointBpl}{\progPointBpl'}
} \\[3mm]
& A_1 \eqdef{} \vaccpred{\texttt{x}}{\texttt{f}}{\texttt{q}}\; A_2 \eqdef{} \vfieldacc{\texttt{y}}{\texttt{g}} > \vfieldacc{\texttt{x}}{\texttt{f}}\;\;
R \eqdef{} \lambda (\stateVpr,\stateBpl).\; \stateRelInst{\ctxtBpl}{\trRecord}{\auxVars}{\stateVpr}{\stateVpr}{\stateBpl} 
\;
R_3 \eqdef{} \lambda (\_,\stateVpr)\;\stateBpl.\; R(\stateVpr, \stateBpl)\; \\
& R_1 \eqdef{} \lambda (\stateVpr^0,\stateVpr)\;\stateBpl.\; \stateVpr^0 = \stateVpr \wedge R(\stateVpr, \stateBpl)  \quad 
R_2 \eqdef{} \stateRelInstBase{\ctxtBpl}{\trRecord_1}{\auxVars} \quad \trRecord_1 \eqdef{} \trRecord(\mathit{M}^0 \mapsto \texttt{WM})
\end{align*}
\caption{
Proof tree constructed by our proof automation for the simulation of $\vexhale{\vsep{\vaccpred{\texttt{x}}{\texttt{f}}{\texttt{q}}}{\vfieldacc{\texttt{y}}{\texttt{g}} > \vfieldacc{\texttt{x}}{\texttt{f}}}}$ via the Boogie statement in \figref{fig:background_vpr_to_bpl_example} on lines \ref{line:background_exh_start}-\ref{line:background_exh_end}.
The automation uses generated hints for the application of rule \stmtSimExhRuleName{}, and
for proofs at the leaves ($\mathcal{P}_i$; left abstract here). 
The Boogie program points $\gamma$, $\gamma_1$, $\gamma_2$, $\gamma_3$, and $\gamma'$ are the points in \figref{fig:background_vpr_to_bpl_example} starting on lines \ref{line:background_exh_start}, \ref{line:background_exh_acc_nonfail_check_start}, \ref{line:background_exh_pure_start}, \ref{line:background_exh_non_det_select}, and \ref{line:background_exh_end}, respectively.
\stateRelInstName{} is our state relation instantiation introduced in~\secref{subsec:state_relation_impl}.
$\mathit{Succ}_2$ is defined in~\figref{fig:stmt_exh_simulation} (where the assertion is $\vsep{A_1}{A_2}$). Rules \exhAuxPropRuleName{} and \exhAuxNoInvSepRuleName{} are derived from \propRuleName{} (\figref{fig:generic_forward_simulation_rules}) and \exhAuxInvSepRuleName{} (\figref{fig:assertion_invariant}), respectively.
}
\label{fig:generated_proof_example}
\end{figure}

We have extended the Viper-to-Boogie implementation to automatically generate an Isabelle proof relating the Viper and Boogie programs.
To make this automatic generation possible, we instrument less than 500 lines of the existing implementation to produce \emph{hints}, which provide extra information about the Boogie encoding.
A core component of our proof automation is an Isabelle tactic that uses these hints to automatically prove forward simulations.
The tactic applies the rules provided by our methodology (\secref{sec:methodology}) to decompose simulations into smaller ones and generates proofs for \emph{atomic simulations} that are not further decomposed.
Our instrumentation generates two kinds of hints for the tactic: (1) hints indicating which candidate of multiple diverse translations is used, and (2) hints specifying how to instantiate parameters and discharge premises of a rule.

As a concrete example, consider~\figref{fig:generated_proof_example}, which shows the proof generated by our tactic (represented via a proof tree) for the forward simulation of $\vexhale{\vsep{\vaccpred{\texttt{x}}{\texttt{f}}{\texttt{q}}}{\vfieldacc{\texttt{y}}{\texttt{g}} > \vfieldacc{\texttt{x}}{\texttt{f}}}}$ via the Boogie statement in \figref{fig:background_vpr_to_bpl_example} on lines \ref{line:background_exh_start}-\ref{line:background_exh_end}.
Hints 1 and 4 in~\figref{fig:generated_proof_example} are hints of the first kind. 
Hint 1 specifies that well-definedness checks are not omitted in the translation of \vexhaleAuxNoArg{}; as a result, the tactic applies \stmtSimExhRuleName{}, which does not track a separate predicate $Q$ on assertions for the \vexhaleAuxNoArg{} simulation (see~\secref{subsec:properties_in_state_relation}).
Hint 4 specifies that the nondeterministic heap assignment is not omitted in the Boogie code (see~\secref{subsec:modularity_and_abstraction} for when it is omitted), which directs the tactic to use a specific rule (not shown in the figure).
Hints 2, 3, and 5 in~\figref{fig:generated_proof_example} are hints of the second kind.
Each of them provides information on temporary Boogie variables (name and lemma showing the declared type is the expected one) in~\figref{fig:background_vpr_to_bpl_example}.
The temporary variables here are (1) \texttt{WM} to set up the expression evaluation state on line~\ref{line:background_exh_start} (hint 2), which results in a change of the translation record in $R_2$ (see~\secref{subsec:state_relation_impl}), (2)~\texttt{tmp} to store the permission on line~\ref{line:background_exh_acc_nonfail_check_start} (hint 3), which is used to adjust the auxiliary variable map (see~\secref{subsec:state_relation_impl}) in proof $\mathcal{P}_2$, and (3) 
\texttt{H'} to perform the nondeterministic selection on line~\ref{line:background_exh_non_det_select} (hint 5).

After decomposing the simulation, our tactic must automatically prove the atomic simulations. In~\figref{fig:generated_proof_example}, $\mathcal{P}_1$ and $\mathcal{P}_4$ are such proofs. $\mathcal{P}_2$ and $\mathcal{P}_3$ further decompose the simulation before reaching atomic simulations ($\mathcal{P}_2$ does so via the rule shown in~\appreftr{app:simulation_rule_example}{B}).
We use two main automation approaches for atomic simulations.
Firstly, we prove (once and for all) simple lemmas about the behaviours of small sequences of simple Boogie commands; these are applied (and their hypotheses discharged) automatically when needed. These lemmas are used for only small parts of the overall translation. Secondly, we prove (once and for all) lemmas that capture effects simulated by \bassumeNoArg{} and \bassertNoArg{} statements for arbitrary expressions.
This generality enables a tactic to automatically prove Viper effects that are simulated via a combination of these two statements.

Our tactic uses both of these approaches for the example in~\figref{fig:generated_proof_example}.
Proof $\mathcal{P}_2$ uses the second approach for justifying the nonfailure check for $\vexhaleAux{\vaccpred{e}{f}{p}}$ shown on lines \ref{line:background_exh_acc_nonfail_check_start}-\ref{line:background_exh_acc_nonfail_check_end} in \figref{fig:background_vpr_to_bpl_example}.\footnote{The approach is designed to work without any changes to the tactic even if the expressions in the two \bassertNoArg{} statements were changed to be syntactically different.}
Proofs $\mathcal{P}_1$ and $\mathcal{P}_4$ use the first approach. 
As part of proof $\mathcal{P}_4$, we use a lemma of the following form proved once and for all ($\mathcal{K}$ is a continuation and the free variables are universally quantified):
\begin{lemma}
If (1) $\stateRelInst{\ctxtBpl}{\trRecord}{\auxVars}{\stateVpr}{\stateVpr}{\stateBpl}$, (2) $\exhNonDetSelect{\stateVpr^0}{\stateVpr}{\stateVpr'}$, (3) $h = \heapTr{\trRecord} \wedge m = \maskTr{\trRecord}$, and (4) ... then there is a state $\stateBpl'$ such that \redStmtBpl{\ctxtBpl}{(\bhavoc{h'} :: \bassume{f(h,h',m)} :: h := h' :: \vec{c}; \mathit{if}, \mathcal{K})}{\normalState{\stateBpl}}{(\vec{c};\mathit{if}, \mathcal{K})}{\normalState{\stateBpl'}} and
$\stateRelInst{\ctxtBpl}{\trRecord}{\auxVars}{\stateVpr'}{\stateVpr'}{\stateBpl'}$.
\end{lemma}
This lemma captures that a \bhavocNoArg{}-\bassumeNoArg{}-assignment sequence simulates the nondeterministic heap assignment \wrt{} our state relation instantiation (see~\secref{subsec:state_relation_impl}).
The fourth premise (not shown here) includes constraints on $f$'s interpretation and on $h'$.\footnote{If the implementation changed the translation for the nondeterministic heap assignment, then we would have to adjust only the tactic's proof strategy for this assignment via a new lemma (\ie{} proof $\mathcal{P}_4$ in~\figref{fig:generated_proof_example}); the rest remains unchanged.}

A general challenge when the tactic applies the rules from~\secref{sec:methodology} is that the Viper and Boogie ASTs are structured differently (see~\secref{subsec:viper_and_boogie_overview}).
Thus, the automatic selection of Boogie program points in the premises of rules is not immediate.
For example, when applying rule~\stmtSimExhRuleName{} in~\figref{fig:generated_proof_example}, the tactic cannot easily choose the intermediate program point $\progPointBpl_3$ by inspecting the initial program point $\progPointBpl$.
Instead, the tactic starts proving the first premise with an \emph{existentially quantified} $\progPointBpl_3$. 
Once the proof reaches the goal of proof $\mathcal{P}_1$ (\ie{} the first atomic simulation), it becomes clear how to advance the program point $\progPointBpl$ and, by the end of the proof of the first premise of \stmtSimExhRuleName{}, the choice of $\progPointBpl_3$ becomes clear.
This strategy is enabled by our routine use of \emph{schematic variables} in Isabelle (\emph{evars} in other tools), for postponing the choice of witnesses for existentially-quantified values.

\subsection{Background Theory and Polymorphic Maps}\label{subsec:background_theory}
Boogie does not have any notion of a heap location or a Viper state. 
Such Viper (and other front-end) constructs are translated using particular global declarations in Boogie.
A subset of the Boogie declarations always emitted by the Viper-to-Boogie implementation is given by:
\begin{itemize}
    \item Uninterpreted types \bcode{bref} and \bcode{bfield} to model references and fields. \bcode{bfield} takes one type argument indicating the type of the corresponding Viper field.\footnote{In practice, \bcode{bfield} takes one more type argument that we ignore for the sake of presentation.}
    \item An uninterpreted function \bcode|GoodMask| that maps a permission map to a Boolean and an axiom restricting this function to return true only if the permission map models a consistent Viper permission mask.
    \item Global variables \bcode|H| and \bcode|M| to model the heap and permission mask, respectively. 
    \bcode|H[x,f]| stores the heap value for heap location \bcode|x.f| and \bcode|M[x,f]| stores the permission value for \bcode|x.f|.
    The types of both variables are represented via Boogie's \emph{impredicatively-polymorphic maps}~\cite{LeinoRPoly10}, which we explain below.
\end{itemize}
\begin{figure*}
\begin{align*}
&\correctProcBpl{G}{p} \eqdef{} \forall \mathcal{T}, \mathcal{F}, \stateBpl.\; [\declsWellDefBpl{G}{p}{\mathcal{T}}{\mathcal{F}} \wedge 
\axiomsSatBpl{G}{\mathcal{T}}{\mathcal{F}}{\stateBpl}] \Longrightarrow \\
&\qquad \qquad \qquad \qquad \forall \progPointBpl', \stateExtBpl'.\; \redStmtBpl{\initCtxtBpl{G}{p}{\mathcal{T}}{\mathcal{F}}}{\initProgPoint{p}}{\normalState{\stateBpl}}{\progPointBpl'}{\stateExtBpl'} \Rightarrow  \stateExtBpl' \neq \failureState{} \\[0.5em]
&\correctMethodVpr{F}{M}{m} \eqdef{} \\
&\quad \forall \stateVpr. \; (\forall l. \mask{\stateVpr}(l) = 0) \Longrightarrow \\
&\qquad \; \forall \stateExtVpr.\; \redStmtVpr{\initCtxtVpr{M}{F}{m}}{\vinhale{\precondVpr{m}};\bodyVpr{m};\vexhale{\postcondVpr{m}}}{\stateVpr}{\stateExtVpr} \Rightarrow \stateExtVpr \neq \failureState{}
\end{align*}
\caption{The correctness definitions for a Boogie procedure $p$ (top) and Viper method $m$ (bottom). We ignore the restriction on well-typed states here, but include the restriction in the Isabelle formalisation.}
\label{fig:correctness_defs}
\end{figure*}
The correctness of a Boogie procedure guarantees no failing executions of the procedure's body for \emph{any} interpretation of the uninterpreted types and functions (1) that is well-formed (\eg{} the function interpretation respects the declared function signatures), and (2) for which all the Boogie axioms in the Boogie program are satisfied in the initial Boogie state.
The formal correctness definition for a Boogie procedure $p$ reflects this directly (a simplified version is shown at the top of~\figref{fig:correctness_defs}). $\mathcal{T}$ and $\mathcal{F}$ are the interpretations of uninterpreted types and functions, respectively. 
$G$ denotes the global declarations in the Boogie program.
$\initProgPoint{p}$ is the initial Boogie program point in the procedure $p$.
$\initCtxtBpl{G}{p}{\mathcal{T}}{\mathcal{F}}$ constructs a Boogie context from the provided parameters.
Thus, to use the correctness of a Boogie procedure, we must choose a type and function interpretation that satisfy the required conditions. 
The main challenge here is formally expressing interpretations that deal with polymorphic Boogie maps, as we discuss next.

\paragraph{Polymorphic maps}
The heap and permission maps are represented (via the Viper-to-Boogie translation) using Boogie's polymorphic maps; this choice is not unusual (\eg{} the Dafny-to-Boogie implementation also currently uses polymorphic maps with similar polymorphic map types as the ones used by the Viper-to-Boogie implementation).
The Boogie maps used to model Viper heaps have the polymorphic map type \texttt{<T>[\texttt{bref, bfield T}]T}: a total map storing, for \emph{any} type \texttt{T}, values of type \texttt{T} given (as key) a reference and field with type argument \texttt{T}.

To our knowledge, there exists no formal model for Boogie's polymorphic maps. Providing a general model is challenging: in particular, Boogie's polymorphic maps are \emph{impredicative}: a map $m$ of type \texttt{<T>[T]T'} permits \emph{any} value as a key, including the map $m$ itself!
Instead of providing a formal model for polymorphic maps in general, we provide one tailored to those that the Viper-to-Boogie implementation uses.
To aid the incorporation of our model, we adjust the implementation to represent a polymorphic map via an uninterpreted type (\eg{} \texttt{HType} for the heap), polymorphic functions for reading from and updating a polymorphic map (\eg{} \texttt{read} and \texttt{upd}), and two axioms that express the relationship between the two functions.
The only change in the translation itself is to simply rewrite heap and mask lookups and updates into calls to these functions; everything else remains identical. 
Then, we provide interpretations of the types and functions, and automatically prove that the axioms hold for these interpretations for any state that maps constants to their defined values; the same approach could be used for \eg{} the Dafny-to-Boogie translation.

%
What remains for our simulation proofs is to provide interpretations for these new components (\eg{} \texttt{HType}, \texttt{read}, and \texttt{upd} for the heap) such that the axioms are fulfilled.
The challenge here is avoiding circularities: \eg{} if the field provided to \texttt{read} has type parameter \texttt{HType}, then the instantiation of \texttt{read} must itself return a heap; to construct an initial heap, we already need a heap of the same type.
To break this circularity, we instantiate \texttt{HType} as a \emph{partial} mapping from reference and fields to values, and allow the empty map to be of type \texttt{HType}, which provides us with a concrete heap.
\texttt{read} is defined to return a default value for reference and field pairs that are not in the domain of the partial map; for heaps the default value is the empty map.
This is sufficient to prove the axioms, since in practice the axioms only require \texttt{read} returning specific values when those values were previously inserted by \texttt{upd}.

\subsection{Generating A Proof of the Final Theorem}\label{subsec:proof_strategy}
We will now discuss, given a Viper program and its Boogie translation, how forward simulation proofs can be used to generate a proof of the final theorem justifying the soundness of the translation: The correctness of the Boogie program (\ie{} the correctness of all contained Boogie procedures) implies the correctness of the Viper program (\ie{} the correctness of all contained Viper methods).

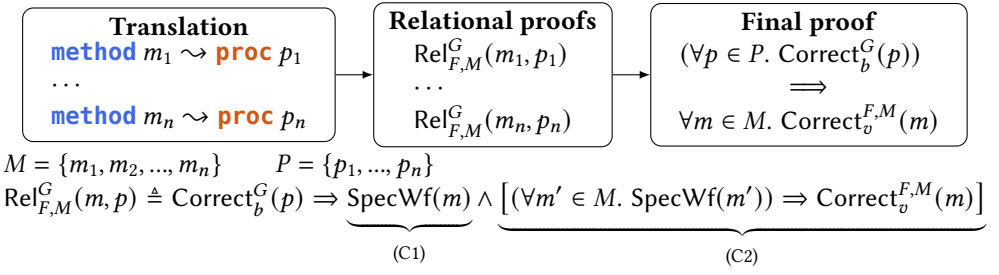
\begin{figure}
  \begin{minipage}{\textwidth}
  \tikzset{>=latex} 
  \centering
  \begin{tikzpicture}
      [statenode/.style = {rectangle, draw, align=center, rounded corners,
              inner xsep=2mm, inner ysep=1mm},
      -latex,     
      ]  
      
      \node[statenode] (pending) {
          \textbf{Translation} \\
            \begin{tabular}{l}
            \vcode|method| $m_1 \translatesto$ \bcode|proc| $p_1$ \\
            $\cdots$ \\
            \vcode|method| $m_n \translatesto$ \bcode|proc| $p_n$ \\
            \end{tabular}
          };

        \node[statenode, right=0.7mm and 5mm of pending.east] (accepted){
            \textbf{Relational proofs} \\
                \begin{tabular}{l}
                \relMethodProc{F}{M}{G}{m_1}{p_1} \\
                $\cdots$ \\
                \relMethodProc{F}{M}{G}{m_n}{p_n} \\
                \end{tabular}
        };

        \node[statenode, right=0.7mm and 5mm of accepted.east] (done){
            \textbf{Final proof} \\
            \begin{tabular}{l}
                $(\forall p \in P.\; \correctProcBpl{G}{p})$ \\
                \multicolumn{1}{c}{$\Longrightarrow$} \\
                $\forall m \in M.\; \correctMethodVpr{F}{M}{m}$
            \end{tabular}
        };



        \draw (pending.east) to (accepted.west);
        \draw (accepted.east) to (done.west);      
  \end{tikzpicture}
    \end{minipage}

    \begin{tabular}{l}
    $M = \{m_1,m_2,...,m_n\} \qquad P = \{p_1,...,p_n\}$ \\
    $\relMethodProc{F}{M}{G}{m}{p} \eqdef \correctProcBpl{G}{p} \Rightarrow  
        \underbrace{\specWellDefVpr{m}}_{\text{(C1)}} \wedge
        \underbrace{\left[(\forall m' \in M.\; \specWellDefVpr{m'})
    \Rightarrow
        \correctMethodVpr{F}{M}{m}
        \right]}_{\text{(C2)}}$
    \end{tabular}
  \caption{Proof strategy for the Viper-to-Boogie translation. First, a proof is generated relating each Viper method with the corresponding Boogie procedure. Second, the final proof is deduced. $F$ denotes the Viper fields, $G$ denotes the global constants, variables, Boogie axioms, and functions emitted by the translation.} 
  \label{fig:proof_workflow}
\end{figure}

We decompose the proof of the final theorem into smaller parts.
At a high level, the Viper-to-Boogie translation works as follows.
Let $F$ and $M$ be the set of Viper fields and methods in the Viper program, respectively.
The Viper-to-Boogie translation (1) emits global Boogie declarations $G$ (see~\secref{subsec:background_theory}) and (2) generates a separate Boogie procedure $p(m)$ for every Viper method $m$ in $M$.
The intended relation between $m$ and $p(m)$ is given by $\relMethodProc{F}{M}{G}{m}{p(m)}$ in~\figref{fig:proof_workflow}, which states that the correctness of $p(m)$ \wrt{} $G$ guarantees two things: (C1) the well-formedness of $m$'s specification, and (C2) the correctness of $m$ \wrt{} $F$ and $M$ \emph{if} the specifications of all methods in the Viper program are well-formed.
The reason that the correctness of $m$ is not implied \emph{directly} is due to the optimised translation of method calls (as explained in~\secref{subsec:vpr_to_bpl_method_calls}).

\figref{fig:proof_workflow} shows how we generate the proof of the desired theorem in two steps.
First, for each Viper method $m$ and its translated Boogie procedure $p(m)$, we generate a proof for $\relMethodProc{F}{M}{G}{m}{p(m)}$, explained next.
Second, we obtain the desired theorem directly from these per-method relational proofs, since the correctness of all Boogie procedures implies that all Viper method specifications are well-formed using (C1), which implies that each Viper method is correct using (C2).

Next, we turn the focus to our strategy for proving $\relMethodProc{F}{M}{G}{m}{p(m)}$. 
For the sake of presentation, we focus on the proof of (C2) (correctness of $m$), and omit the proof of (C1) (well-formedness of $m$'s specification).
Intuitively, to prove that $m$ is correct, we have to show that for any state that satisfies $m$'s precondition, executing $m$'s body in this state results in a state that satisfies $m$'s postcondition.
The correctness definition for a Viper method (shown at the bottom of~\figref{fig:correctness_defs}) expresses this by requiring that any execution starting in a state $\stateVpr$ with no permissions that inhales the precondition, then executes the body, and finally exhales the postcondition, cannot fail. 
As planned, we obtain this result via a forward simulation proof between the executed Viper statement and $p(m)$'s procedure body using our presented methodology.
Formally, we show:
\begin{align*}\exists R', \progPointBpl'.\; \stmtSim{\ctxtVpr^0}{\ctxtBpl^0}{R_0}{R'}{s_v^0}{\initProgPoint{p(m)}}{\progPointBpl'} \\
\textit{where } s_v^0 \eqdef{} \vinhale{\precondVpr{m}};\bodyVpr{m};\vexhale{\postcondVpr{m}} 
\end{align*}
In the statement above, $\ctxtVpr^0 \eqdef{} \initCtxtVpr{M}{F}{m}$ is the initial Viper context. $\ctxtBpl^0$ is a Boogie context that is defined in terms of our chosen type and function interpretation (see~\secref{subsec:background_theory}).
$R_0$  is an instantiation of the state relation shown in~\secref{subsec:state_relation_impl}.
$\initProgPoint{p(m)}$ is the initial Boogie program point in $p(m)$.
The output state relation and output Boogie program point are irrelevant, since we care only about the simulation of failing Viper executions here.
To complete the proof, we choose an initial Boogie state $\stateBpl$ such that $R_0(\stateVpr, \stateBpl)$.
As a result, if a Viper execution $E_v$ of statement $s_v^0$ in $\stateVpr$ \emph{fails}, the forward simulation provides us with a failing Boogie execution $E_b$ of $p(m)$.
Using the correctness of $p(m)$, we conclude that $E_b$ cannot fail, and thus conclude that $E_v$ cannot fail, which concludes the proof of $\relMethodProc{F}{M}{G}{m}{p(m)}$.

\section{Implementation and Evaluation}\label{sec:evaluation}

We instrumented the existing Viper verifier implementation to automatically produce an Isabelle proof justifying the soundness of its translation to Boogie, and evaluated this validation on a diverse set of Viper benchmarks.

\paragraph{Implementation.}
Even though Viper passes the generated Boogie program to Boogie as a text file, our soundness proof directly connects the input Viper AST to the internal AST representation of the Boogie verifier. Therefore, we do not have to trust the Boogie parser.

We make the following four adjustments to the Viper verifier implementation.
First, we desugar the uses of polymorphic maps as described in~\secref{subsec:background_theory}, since there is no formal model for polymorphic maps.
Second, we adjust the implementation to not emit Boogie declarations or commands that are used only for features outside of our subset (the implementation always emits those without checking whether the corresponding features are actually used).
Third, we switch off simple syntactic transformations that the Viper verifier applies to the produced Boogie program  (\eg{} constant folding, elimination of if-statements with no branches), since we do not support them yet; justifying those transformations should be straightforward and is orthogonal to our work.
Fourth, we introduce a \bhavocNoArg{} statement in the Boogie program at the point when a scoped Viper variable is introduced, which faithfully models the semantics of such a variable.
The original Viper implementation instead just introduces a fresh Boogie variable at the beginning of the Boogie procedure. Proving the equivalence of both translations is straightforward.

{
\small
\begin{table}
	\caption{Overview of benchmarks and results. For each test suite, we report the number of Viper files, the total number of Viper methods contained in those files, as well as the \emph{mean} number of non-empty lines of code for the Viper files, Boogie files, and produced Isabelle proofs. We measured the mean and median time it took to check the Isabelle proofs in seconds.}
	\begin{center}
		\begin{tabular}{lrrrrrrr}
			\toprule
			\textbf{Test suite} & \textbf{Files} & \textbf{Methods} & \textbf{Viper} & \textbf{Boogie} & \textbf{Isabelle} & \multicolumn{2}{c}{\textbf{Proof Check}}   \\
& no. & no. & Mean [LoC] & Mean [LoC] & Mean [LoC] & Mean [s] & Median [s] \\
			\midrule
			Viper & 34 & 105 & 33 & 298 & 1719 & ~33.8 & ~23.8 \\
			Gobra & 17 & 65 & 60 & 287 & 1937 & ~32.7 & ~25.3 \\
			VerCors & 18 & 116 & 63 & 332 & 2930 & ~43.1 & ~40.9 \\
			MPP & 3 & 13 & 206 & 1060 & 5164 & ~109.0 & ~46.2 \\
			\textbf{\textbf{Overall}} & \textbf{72} & \textbf{299} & \textbf{54} & \textbf{335} & \textbf{2217} & \textbf{~39.0} & \textbf{~32.9} \\
			\bottomrule
		\end{tabular}
	\end{center}
	\label{tbl:results}
\end{table}

}

\paragraph{Evaluation.}
Our evaluation answers the questions: \textbf{(RQ1)} Does our implementation generate proofs that Isabelle can check successfully for a diverse set of examples? \textbf{(RQ2)} Does Isabelle check the generated proofs in reasonable time (\eg{} feasible as part of continuous integration)?

To obtain a diverse set of representative examples,
we considered the Viper test suite as well as the test suites of three tools that produce Viper code:
Gobra~\cite{WolfACOPM21} (for  Go),
VerCors~\cite{BlomHuisman14} (for Java),
and MPP~\cite{EilersMH18} (a tool performing a modular product transformation on Viper programs).
To eliminate trivial translations, we focused on Viper programs that use the heap, as indicated by the occurrence of at least one accessibility predicate.
Out of those, we included all Viper programs that fall into our supported Viper subset. We followed different strategies to systematically obtain additional examples from the different test suites. 
For Viper and MPP, we additionally included all files that have an \emph{old}-expression (by manually removing the corresponding assertion, \ie{} verifying weaker postconditions)
or a \emph{new} statement (by manually desugaring the allocation primitive into our subset).
Moreover, we made sure that each argument to a method call is a variable (e.g. we rewrote \vcode{m(i+1)} to \vcode{var t := i+1; m(t)}), since we currently support only variables as arguments.
For Gobra and VerCors, we removed boilerplate code that is emitted for each file and then followed the same process as for Viper and MPP\@.
Moreover, we additionally included files generated by Gobra that had at most two occurrences of features outside of our subset if those 
could be desugared into our subset (\eg{} by eliminating a function by inlining its body).

As summarised in \tabref{tbl:results}, we collected a total of 72 Viper files (containing 299 methods), with a mean of 54 and maximum of 414 non-empty lines of code.
We ran our implementation on all 72 Viper files to generate the Boogie translations and the Isabelle proofs, and measured the time it took for Isabelle to check the generated proofs (the mean of five repetitions). 
The measurements were run on a ThinkPad X1 Yoga Gen 5 on Ubuntu 20.04 with 16 GB RAM and i7-10510U @ 1.8 GHz (scaled up to 4.9 GHz using Turbo Boost).
The generated Boogie translations are on average 6.2x larger (335 non-empty LoC on average), illustrating the semantic gap between Viper and Boogie.

Isabelle successfully checked the generated proofs for all Viper files,
including the Viper programs automatically generated by other tools. This shows that our approach is effective for practical verifiers and answers RQ1 positively. 
The resulting Isabelle proofs have on average over 2000 lines and are checked in less than a minute. 

{
\footnotesize
\begin{table}
	\caption{Detailed results of our evaluation for a selection of files showing the number of methods, the nonempty lines of code for the Viper program, Boogie program, and produced Isabelle proof, and the time it took to check the proof in seconds.}
	\begin{center}
		\begin{tabular}{llrrrrr}
			\toprule
			\textbf{Test suite} & \textbf{File} & \textbf{Methods} & \textbf{Viper} & \textbf{Boogie} & \textbf{Isabelle} & \textbf{Proof Check} \\
& & no. & Total [LoC] & Total [LoC] & Total [LoC] & Total [s] \\
			\midrule
			Viper & testHistoryProcesses~ & 13 & 205 & 1711 & 7035 & ~126.3 \\
			Gobra & defer-simple-02~ & 9 & 211 & 853 & 4717 & ~60.6 \\
			VerCors & inv-test-fail2~ & 5 & 92 & 514 & 2596 & ~56.5 \\
			MPP & banerjee~ & 8 & 414 & 2014 & 9545 & ~242.4 \\
			MPP & darvas~ & 2 & 91 & 582 & 2800 & ~38.4 \\
			MPP & kusters~ & 3 & 112 & 583 & 3146 & ~46.2 \\
			\bottomrule
		\end{tabular}
	\end{center}
	\label{tbl:selection}
\end{table}

}

\tabref{tbl:selection} shows the results for a selection of examples (the detailed results for each test suite are shown in~\appreftr{app:results_evaluation}{D}):
All three examples from MPP, as well as the largest example (in terms of lines of Viper code) from each of the other test suites.
The three MPP examples are drawn from different research papers and show that our tool can certify challenging programs.

For this selection, the times to check the proofs range from 38 seconds to 4 minutes.
No file in any of the 72 files takes longer than 4 minutes to check.
These times are acceptable, since we expect the validation to be performed occasionally (in particular, before the verified program is released or as part of continuous integration), but not on every run of the verifier. 
Thus, we answer RQ2 positively for the considered 72 files. To obtain additional representative files from the test suites, we would need to extend the supported Viper subset.
Moreover, most of our proofs are not yet optimised to make proof checking faster.
For example, field and variable accesses currently result in an overhead in the proof that is proportional to the number of fields and active variables, respectively.
This could be improved by constructing and updating lookup tables efficiently.

\section{Related Work}\label{sec:related_work}

Various works prove the soundness of front-end translations  \emph{once and for all}. For instance, \citet{LehnerMueller2007}
prove a simplified translation from Java Bytecode to Boogie, and 
\citet{VogelsJP09} target a translation from a toy object-oriented programming language to Boogie. Both proofs are done on paper and do not consider an actual implementation of the translation.
\citet{Backes2011} prove a translation sound from the Dminor data processing language to the Bemol IVL in Coq.
They do not provide a proof connecting the formalised translation to their F{\#} implementation.
\citet{Herms13} proves a translation from C to the WhyCert IVL (inspired by the Why3 IVL) sound in Coq, which they then turn into an executable tool via Coq's extraction to OCaml. The resulting tool has similarities to the Jessie Frama-C implementation~\cite{JessieFramaCPlugin}, which translates C programs to Why3; \citet{Herms13} discusses mismatches between their mechanisation and the Jessie implementation. In contrast, our certification applies to existing front-end implementations, which are typically implemented in efficient mainstream programming languages, use diverse libraries, and include subtle optimisations omitted from idealised implementations.
\citet{SmansJacobsPiessens12} prove soundness of a verification condition generator for a language with implicit dynamic frames (IDF) assertions once and for all on paper without using an IVL. They also implement a prototype, but do not formally connect the proof to the implementation. We also applied our methodology to a verifier based on IDF, but validate an actual implementation.

Many verifiers perform a series of program transformations, \eg{} by translating programs to a lower-level IVL or internally without changing the language (e.g. monomorphisation). 
Our approach can in principle be applied to both kinds of transformations, but is tailored towards the former, where the semantic gap is large, non-local checks arise, and diverse translations are used.
Existing work for the validation of internal transformations does not provide solutions for these challenges.
For instance, our prior work~\cite{ParthasarathyMuellerSummers21} validates the verification condition (VC) generation implementation of Boogie programs (also via a proof-producing instrumentation), which includes internal Boogie-to-Boogie transformations.
In these transformations, the semantic gap is small (the source and target constructs are largely the same), and thus the decomposition into smaller problems is immediate, while in this paper the decomposition is a challenge.
Moreover, our prior work need not deal with non-local checks and diverse translations.
Our prior work uses different kinds of simulations; it would be interesting future work to apply our methodology to these.
Besides internal transformations, our prior work connects the VC and a Boogie program; this paper considers only program-to-program transformations.
Our prior work can in principle be combined with this paper to enable end-to-end soundness guarantees for Viper, but requires extending the Boogie verifier validation to more internal Boogie transformations and to a larger Boogie subset.

Validation has also been used to obtain formal guarantees for implementations of other verifiers, but none of the existing works target front-end translations and the challenges they entail.
\citet{LinCTWR23} and~\citet{WilsJacobs2023} validate verifiers obtained via the K framework and VeriFast, respectively. These verifiers use symbolic execution, which requires a fundamentally different validation approach. \citet{Garchery21} validates certain logical transformations in Why3.
\citet{Cohen24} also prove such logical transformations, but do so once and for all in Coq to demonstrate their Why3 mechanisation.
Neither of the two consider the actual VC generation.

Multiple works also embed programs in an interactive theorem prover (ITP) and then automate forward simulation proofs.
\citet{RizkallahLNSCOM16} define a refinement calculus for the Cogent compiler to automatically prove a forward simulation in Isabelle for a Cogent expression and its C translation.
Their calculus includes syntax-directed rules for deriving simulation judgements, but these rules do not provide the abstraction we needed to handle diverse translations. 
The compiler  was developed with validation in mind, which simplifies, for instance, the treatment of optimisations. In contrast, our goal was to validate existing verifier implementations with all their intricacies.
The verification of the seL4 kernel includes two large forward simulation proofs in Isabelle, for which proof automation was developed~\cite{WinwoodKSACN09,CockKS08,KleinSW10}. This automation reduces the manual proof overhead, but still requires user interaction. In contrast, our validation proofs are generated and checked completely automatically. 
Like us, they prove rules to decompose the simulation for composite statements syntactically but, contrary to us, do not decompose statements semantically into smaller simulations.
They turn certain simulation judgements into Hoare triples for which they have separate automation. 

\emph{Formal translation validation} approaches for compilers express a per-run validator in an ITP~\cite{TristanL08,TristanL09,GourdinBBMB23}, prove it correct once and for all, and then extract executable code (the extraction must be trusted).
For many of these validators, the source and target languages are similar.
It would be interesting to test the feasibility of such approaches for front-end translations, where the semantic gap between the languages is large.
Another difference is that front-end translations incorporate reasoning steps, such as assumptions and proof obligations prescribed by a program logic. This encoding is achieved via components not present in executable languages such as assume statements, havoc statements, and axiomatisations. Moreover, front-end translations emit code that checks nontrivial properties that are then relied upon in other parts of the encoding.

\citet{ZimmermanDA24} define a formal Viper semantics for a Viper subset in order to prove formal results for the gradual verifier Gradual C0 that uses Viper. However, in contrast to ours, their formalisation is not mechanised.
Boogie developers have added an option to monomorphise polymorphic maps in Boogie programs via non-polymorphic maps~\cite{BoogieMonomorphisationPR}. 
This option provides an alternative to ours for desugaring polymorphic maps, which, in the case of Viper, circumvents the circularity challenge discussed in~\secref{subsec:background_theory}, since Viper does not permit storing heaps in fields.
However, in general, front-ends may permit storing heaps in fields.

\section{Conclusion}\label{sec:conclusion}

We presented a methodology for the validation of the front-end translations implemented in practical automated program verifiers. We demonstrated that it handles the complexity and intricacies of the Viper-to-Boogie translation as implemented in the Viper tool. To the best of our knowledge, this is the first formal soundness guarantee for a practical front-end translation. Together with existing work on back-end (and SMT) validation, our work provides a path towards trustworthy automated verifiers.
Two fundamental requirements of our approach are the existence of a formal semantics for the input language and IVL, and the ability to instrument the verifier implementation.
As future work, we plan to extend the supported Viper subset and to apply our methodology to verifiers that target Viper as an IVL and that verify, for instance, concurrent or object-oriented programs.

\begin{acks}
We thank Aleksandar Hubanov for work on embedding the Boogie AST in Isabelle, Marco Eilers for helping with the modular product program tool, Jo{\~{a}}o C. Pereira and Felix A. Wolf for helping with the Gobra verifier, Xavier Denis for clarifications on Why3 and Michael Sammler for feedback on our formalisation. We thank the anonymous reviewers for their comments. 
This work was partially funded by the Swiss National Science Foundation (SNSF) under Grant No. 197065.
\end{acks}

\section*{Data Availability Statement}
Our publicly-available artifact~\cite{artifact} contains:
\begin{enumerate}
  \item an Isabelle formalisation for the technical results in \secref{sec:background}, \secref{sec:methodology}, and \secref{sec:validation_impl}.
  \item our proof-producing Viper-to-Boogie implementation, which generates, on every run of the verifier, an Isabelle proof showing that the correctness of the input Viper program is implied by the correctness of the corresponding Boogie translation.
  \item the examples used in the evaluation, and scripts for the benchmark selection and evaluation results (\secref{sec:evaluation}).
\end{enumerate}

\bibliography{references}

\ifx\istr\undefined
\end{document}
\else
\clearpage
\appendix
\section{Inhale Semantics}\label{app:inhale_semantics}
\begin{figure}
\begin{align*}
%
&\Inf[\rulename{inh}]
{\redInh{A}{\stateVpr}{\stateExtVpr}}.
{\redStmtVpr{\ctxtVpr}{\vinhale{A}}{\stateVpr}{\stateExtVpr}}
%
&\Inf[\rulename{inh-acc}]
{\redExprVpr{e}{\stateVpr}{\normalVal{r}}}
{\redExprVpr{e_p}{\stateVpr}{\normalVal{p}}}.
{p < 0 \Rightarrow \stateExtVpr = \failureState{}}.
{ p \geq 0 \Rightarrow \stateExtVpr = \textit{if}\; \mathit{inhSucc(r,p)}\; \textit{then}\; \normalState{\stateVpr'} \; \textit{else} \; \magicState{}}.
{\stateVpr' = \textsf{addperm}(\stateVpr, r, f, p)}
{\redInh{\vaccpred{e}{f}{e_p}}{\stateVpr}{\stateExtVpr}} \\[0.6em]
%
%
& \Inf[\rulename{inh-sep-s}]{\redInh{A}{\stateVpr}{\normalState{\stateVpr'}}}.
{\redInh{B}{\stateVpr'}{\stateExtVpr}}.
{\redInh{A*B}{\stateVpr}{\stateExtVpr}}
%
%
& \Inf[\rulename{inh-sep-f}]{\redInh{A}{\stateVpr}{\failureState{}}}.
{\redInh{A*B}{\stateVpr}{\failureState{}}} \\[3mm]
\end{align*}
$\textit{inhSucc}(r,p) \eqdef{} (p > 0 \Rightarrow r \neq \vnull{}) \wedge{} (r \neq \vnull{} \Rightarrow p+\mask{\stateVpr}(r,f) \leq 1)$
\caption{A subset of the rules for the formal semantics of inhale. $\textsf{addperm}(\stateVpr, r, f, p)$ denotes the state $\stateVpr$ where permission $p$ has been added to $(r,f)$.}
\label{fig:semantics_inhale}
\end{figure}

The reduction of $\vinhale{A}$ for an assertion $A$ from state $\stateVpr$ to outcome $\stateExtVpr$ is expressed via the judgement $\redInh{A}{\stateVpr}{\stateExtVpr}$.
The rule~\rulename{inh} in~\figref{fig:semantics_inhale} converts such a judgement into the standard statement reduction judgement.
The rules for the separating conjunction and the accessibility predicate (when the receiver and permission are well-defined) are shown in~\figref{fig:semantics_inhale}.
In the case of the accessibility predicate, there is an additional rule where the \vinhaleNoArg{} fails if $e$ or $e_p$ are not well-defined, which is not shown in the figure.

The accessibility predicate rule~\rulename{inh-acc} shown in \figref{fig:semantics_inhale} expresses that if the added permission is negative then the operation fails.
If the permission is nonnegative, then the operation succeeds if (1) the receiver is non-null if $p > 0$, and (2) the added permission does not yield an inconsistent state (\ie{} does not result in more than 1 permission for $(r,f)$). 
Otherwise, the operation stops (denoted by outcome \magicState{}).
If the operation succeeds, then the new state additionally contains the added permission $p$ at location $(r,f)$.

%

\section{Another Simulation Rule Example}\label{app:simulation_rule_example}

\begin{figure}
\begin{align*}
&\Inf[\exhAuxAccRuleName]
{\forall \sigma_v.\; \exprWfSim{\ctxtBpl}{\hat{R}(\sigma_v)}{\hat{R_A}(\sigma_v)}{[e,e_p]}{\progPointBpl}{\progPointBpl_1}}
{\text{(subexpression well-definedness)}}.
    { \forall r,p. \;
    \genericSim{\ctxtBpl}{R_A}{R_B(r,p)}{\mathit{Succ}_A(r,p)}{\mathit{Fail}_A(r,p)}{\progPointBpl_1}{\progPointBpl_2}}
    {\text{(non-failure check)}}
    .
{ \forall r,p. \;
      \genericSim{\ctxtBpl}{R_B(r,p)}{R'}{\mathit{Succ}_B(r,p)}{(\lambda \_.\; \bot)}{\progPointBpl_2}{\progPointBpl'}
}
{\text{(state update)}}
{\exhAuxSim{\ctxtBpl}{R}{R'}{\vaccpred{e}{f}{e_p}}{\progPointBpl}{\progPointBpl'}} \\[3mm]
& \hat{R}(\stateVpr) \eqdef{} \lambda \stateVpr^0\;\stateBpl.\; R((\stateVpr^0,\stateVpr),\stateBpl) \quad 
\hat{R_A}(\stateVpr) \eqdef{} \lambda \stateVpr^0\;\stateBpl.\; R_A((\stateVpr^0,\stateVpr),\stateBpl) \\
& \mathit{Succ}_A(r,p) \eqdef{} 
   \left(
  \lambda (\stateVpr^0,\stateVpr)\;(\stateVpr^1,\stateVpr').\;
  \begin{array}{l}
   {\color{exhAccSuccColor} \exhAccSucc{r}{p}{\stateVpr}} \wedge{}
   (\stateVpr^0,\stateVpr) = (\stateVpr^1,\stateVpr') \wedge{} \\ 
   {\color{wfAccAssmsColor} \wfAccAssms{e}{e_p}{r}{p}{\stateVpr^0}}
  \end{array}
  \right) \\
& \mathit{Fail}_A(r,p) \eqdef{} \lambda (\stateVpr^0,\stateVpr).\;  
\neg{\color{exhAccSuccColor} \exhAccSucc{r}{p}{\stateVpr}} \wedge{}
    {\color{wfAccAssmsColor} \wfAccAssms{e}{e_p}{r}{p}{\stateVpr^0}}
     \\
& \mathit{Succ}_B(r,p) \eqdef{}
  \left(
  \lambda (\stateVpr^0,\stateVpr)\;(\stateVpr^1,\stateVpr').\; 
  \begin{array}{l}
             \stateVpr' = \removePerm{\stateVpr}{r}{f}{p} \wedge{}
             \stateVpr^0 = \stateVpr^1 \wedge{} \\
             {\color{exhAccSuccColor} \exhAccSucc{r}{p}{\stateVpr}} \wedge{} 
             {\color{wfAccAssmsColor} \wfAccAssms{e}{e_p}{r}{p}{\stateVpr^0}}
  \end{array} 
  \right) \\
& {\color{wfAccAssmsColor} \wfAccAssms{e}{e_p}{r}{p}{\stateVpr^0}} \eqdef{} \redExprVpr{e}{\stateVpr^0}{\normalVal{r}} \wedge  \redExprVpr{e_p}{\stateVpr^0}{\normalVal{p}}
\end{align*}
\caption{Rule for the simulation of $\vexhaleAux{\vaccpred{e}{f}{e_p}}$. The definition of {\color{exhAccSuccColor} \exhAccSuccNoArg{}} is given in~\figref{fig:semantics_exhale}.}
\label{fig:exh_aux_acc_simulation}
\end{figure}

Consider the rule~\exhAuxAccRuleName{} in~\figref{fig:exh_aux_acc_simulation}, which decomposes the simulation of~$\vexhaleAux{\vaccpred{e}{f}{e_p}}$ (ignore the universal quantifiers for now)
into the simulation of three separate Viper effects: (1)~the check of well-definedness of the receiver $e$ and permission expression $e_p$ (via the \exprWfSimNoArg{} instantiation from~\figref{fig:forward_simulation}), (2)~a check {\color{exhAccSuccColor} \exhAccSuccNoArg{}} ensuring that the operation will not fail (from the semantics; see~\figref{fig:semantics_exhale}), and (3)~the actual update of the Viper state, which removes the permission.


The second premise includes contextual information, namely the conjunct {\color{wfAccAssmsColor} \wfAccAssmsNoArg{}} expressing that $e$ and $e_p$ are well-defined (which is ensured by the first premise) and evaluate to the reference value $r$ and permission value $p$.
The third premise modelling the removal of the permission includes the same conjunct and that the operation will succeed ({\color{exhAccSuccColor} \exhAccSuccNoArg{}}).
Without the latter, we could in general not prove that the resulting Boogie state satisfies crucial invariants, for instance, that none of the permissions stored in the Boogie state are negative. Again, we are agnostic as to syntactically \emph{how} this is achieved by this check: our rule does \emph{not} require the Boogie program to emit an explicit Boogie \bassertNoArg{} command checking that the permission is nonnegative.
This is important, since the implementation omits such a command, for example, if the permission is the literal 1.

The first universal quantifier is technically motivated: it 
expresses that the simulation must hold for \emph{any} \exhAuxReductionState{}.
The other quantifiers over reference values $r$ and permission values $p$ make the rule more powerful and reusable.
They permit the relation $R_B$ to directly talk about the values that $e$ and $e_p$ evaluate to as specified by the success and failure predicates.
This is particularly useful for justifying cases where the simulation of the non-failure check establishes a property on $r$ or $p$, which is then used in the simulation of the state update.
For example, the Viper-to-Boogie translation stores $p$ into an auxiliary variable that is used for both the non-failure check and the state update.

\section{Instantiation-Independent Consequence Rule}\label{app:conseq_rule}
\begin{figure}
\Inf[\conseqRuleName]
{\genericSim{\ctxtBpl}{R_0'}{R_1'}{S'}{F'}{\progPointBpl}{\progPointBpl'}}.
{\forall \stateVpr\; \stateBpl.\; R_0(\stateVpr, \stateBpl) \Rightarrow R_0'(\stateVpr, \stateBpl) \quad \text{(weaker input state relation)}}.
{\forall \stateVpr\;\stateVpr'.\; S(\stateVpr, \stateVpr') \Rightarrow S'(\stateVpr, \stateVpr') \quad {\text{(weaker success predicate)}}}.
{\forall \stateVpr.\; F(\stateVpr) \Rightarrow F'(\stateVpr) \quad \text{(weaker failure predicate)}}.
{\forall \stateVpr\; \stateVpr'\; \stateBpl'.\; (\exists \stateBpl.\; R_0(\stateVpr, \stateBpl)) \Rightarrow R_1'(\stateVpr', \stateBpl') \Rightarrow S(\stateVpr, \stateVpr') \Rightarrow R_1''(\stateVpr', \stateBpl')}.
{\genericSim{\ctxtBpl}{R_0}{R_1''}{S}{F}{\progPointBpl}{\progPointBpl'}}
\caption{An instantiation-independent consequence rule for the generic forward simulation judgement.}
\label{fig:conseq_rule}
\end{figure}

The rule~\conseqRuleName{} in~\figref{fig:conseq_rule} shows an instantiation-independent consequence rule for the generic forward simulation judgement.
The rule enables proving a simulation judgement by proving a different simulation judgement in the premise where the input state relation, success predicate, and failure predicate are weakened. 
The output state relation is potentially adjusted in the premise as well.
Additionally, the final premise expresses a condition ensuring that the output state relation $R_1''$ in the conclusion will be established for the output states if the initial state relation holds for the initial states, the adjusted output relation holds on the output states, and the Viper execution is successful.

\section{Detailed Results of the Evaluation}\label{app:results_evaluation}
{
\small
\begin{table}
	\caption{Detailed results of our evaluation for the files from the test suite of Gobra.}
	\begin{center}
		\begin{tabular}{llllll}
			\toprule
			\textbf{File} & \textbf{Methods} & \textbf{Viper} & \textbf{Boogie} & \textbf{Isabelle} & \textbf{Proof Check} \\
 & no. & Total [LoC] & Total [LoC] & Total [LoC] & Total [s] \\
			\midrule
			concurrency~ & 2 & 24 & 164 & 1153 & ~25.3 \\
			defer-simple-01~ & 6 & 142 & 639 & 3344 & ~49.6 \\
			defer-simple-02~ & 9 & 211 & 853 & 4717 & ~60.6 \\
			perm-fail1~ & 15 & 165 & 661 & 6392 & ~66.5 \\
			perm-simple1~ & 9 & 131 & 622 & 4221 & ~50.7 \\
			fail1~ & 3 & 44 & 283 & 1574 & ~31.6 \\
			fail3~ & 2 & 19 & 116 & 1044 & ~23.6 \\
			simple1~ & 2 & 30 & 237 & 1210 & ~28.4 \\
			simple2~ & 1 & 10 & 90 & 672 & ~21.4 \\
			simple3~ & 1 & 17 & 186 & 801 & ~24.6 \\
			global-const-8~ & 6 & 49 & 206 & 2510 & ~33.8 \\
			pointer-identity~ & 1 & 30 & 158 & 731 & ~23.0 \\
			pointer-identity~ & 1 & 30 & 158 & 731 & ~23.1 \\
			000008~ & 1 & 10 & 85 & 672 & ~21.4 \\
			000009~ & 1 & 16 & 98 & 679 & ~21.3 \\
			000039~ & 3 & 49 & 178 & 1410 & ~26.8 \\
			000155~ & 2 & 39 & 152 & 1075 & ~24.3 \\
			\bottomrule
		\end{tabular}
	\end{center}
	\label{tbl:Gobra}
\end{table}

\begin{table}
	\caption{Detailed results of our evaluation for the files from the test suite of MPP.}
	\begin{center}
		\begin{tabular}{llllll}
			\toprule
			\textbf{File} & \textbf{Methods} & \textbf{Viper} & \textbf{Boogie} & \textbf{Isabelle} & \textbf{Proof Check} \\
 & no. & Total [LoC] & Total [LoC] & Total [LoC] & Total [s] \\
			\midrule
			banerjee~ & 8 & 414 & 2014 & 9545 & ~242.4 \\
			darvas~ & 2 & 91 & 582 & 2800 & ~38.4 \\
			kusters~ & 3 & 112 & 583 & 3146 & ~46.2 \\
			\bottomrule
		\end{tabular}
	\end{center}
	\label{tbl:MPP}
\end{table}

\begin{table}
	\caption{Detailed results of our evaluation for the files from the test suite of VerCors.}
	\begin{center}
		\begin{tabular}{llllll}
			\toprule
			\textbf{File} & \textbf{Methods} & \textbf{Viper} & \textbf{Boogie} & \textbf{Isabelle} & \textbf{Proof Check} \\
 & no. & Total [LoC] & Total [LoC] & Total [LoC] & Total [s] \\
			\midrule
			BasicAssert-e1~ & 6 & 41 & 197 & 2589 & ~35.0 \\
			BasicAssert~ & 6 & 41 & 193 & 2589 & ~35.0 \\
			DafnyIncr~ & 8 & 60 & 265 & 3419 & ~41.6 \\
			DafnyIncrE1~ & 8 & 57 & 220 & 3340 & ~40.2 \\
			permissions~ & 5 & 39 & 208 & 2270 & ~33.1 \\
			inv-test-fail1~ & 5 & 90 & 510 & 2589 & ~55.5 \\
			inv-test-fail2~ & 5 & 92 & 514 & 2596 & ~56.5 \\
			inv-test~ & 5 & 90 & 510 & 2589 & ~55.1 \\
			SwapIntegerFail~ & 8 & 79 & 429 & 3645 & ~49.8 \\
			SwapIntegerPass~ & 8 & 81 & 469 & 3688 & ~53.0 \\
			SwapLong~ & 6 & 57 & 277 & 2725 & ~36.7 \\
			SwapLongTwice~ & 8 & 81 & 469 & 3688 & ~52.1 \\
			SwapLongWrong~ & 8 & 79 & 429 & 3645 & ~48.9 \\
			frame-error-1~ & 5 & 35 & 173 & 2191 & ~32.8 \\
			refute3~ & 6 & 49 & 246 & 2662 & ~34.5 \\
			refute4~ & 6 & 54 & 258 & 2676 & ~35.9 \\
			refute5~ & 6 & 50 & 253 & 2662 & ~35.9 \\
			demo1~ & 7 & 60 & 347 & 3185 & ~44.6 \\
			\bottomrule
		\end{tabular}
	\end{center}
	\label{tbl:VerCors}
\end{table}

\begin{table}
	\caption{Detailed results of our evaluation for the files from the test suite of Viper.}
	\begin{center}
		\begin{tabular}{llllll}
			\toprule
			\textbf{File} & \textbf{Methods} & \textbf{Viper} & \textbf{Boogie} & \textbf{Isabelle} & \textbf{Proof Check} \\
 & no. & Total [LoC] & Total [LoC] & Total [LoC] & Total [s] \\
			\midrule
			0004~ & 1 & 6 & 100 & 729 & ~21.7 \\
			0004-CPG1~ & 1 & 6 & 95 & 704 & ~21.6 \\
			0005~ & 1 & 4 & 78 & 665 & ~21.1 \\
			0008~ & 2 & 12 & 241 & 1396 & ~26.8 \\
			0011~ & 5 & 63 & 902 & 3284 & ~55.7 \\
			0015~ & 1 & 6 & 92 & 709 & ~21.4 \\
			0052~ & 1 & 7 & 100 & 719 & ~21.5 \\
			0063~ & 6 & 34 & 180 & 2595 & ~35.2 \\
			0072~ & 1 & 8 & 112 & 770 & ~22.4 \\
			0073~ & 1 & 10 & 132 & 737 & ~22.2 \\
			0088-1~ & 1 & 9 & 115 & 751 & ~21.9 \\
			0094~ & 1 & 6 & 91 & 679 & ~21.2 \\
			0152~ & 2 & 14 & 139 & 1137 & ~24.5 \\
			0157~ & 8 & 47 & 354 & 3508 & ~45.1 \\
			0159~ & 2 & 13 & 120 & 1083 & ~23.8 \\
			0170~ & 1 & 8 & 84 & 665 & ~21.1 \\
			0177-1~ & 1 & 10 & 102 & 665 & ~21.4 \\
			0222~ & 2 & 13 & 118 & 1054 & ~23.9 \\
			0227~ & 1 & 5 & 85 & 683 & ~21.4 \\
			0324~ & 1 & 7 & 104 & 704 & ~21.2 \\
			0345~ & 3 & 21 & 165 & 1463 & ~24.4 \\
			0384~ & 1 & 11 & 127 & 709 & ~22.0 \\
			assert~ & 1 & 7 & 92 & 693 & ~21.5 \\
			negative-amounts~ & 3 & 21 & 155 & 1517 & ~27.4 \\
			old~ & 6 & 38 & 318 & 2805 & ~37.9 \\
			swap~ & 2 & 16 & 177 & 1239 & ~25.6 \\
			test~ & 1 & 6 & 81 & 663 & ~20.9 \\
			testHistoryProcesses~ & 13 & 205 & 1711 & 7035 & ~126.3 \\
			testHistoryProcessesPVL~ & 13 & 204 & 1711 & 7035 & ~116.3 \\
			testHistoryProcessesPVL-CPG1~ & 4 & 56 & 490 & 2304 & ~46.1 \\
			testHistoryThreadsProcessesPVL~ & 4 & 56 & 490 & 2304 & ~45.7 \\
			test-example1~ & 4 & 57 & 374 & 2152 & ~37.0 \\
			test-example3~ & 5 & 74 & 430 & 2634 & ~39.3 \\
			test-example4~ & 5 & 71 & 451 & 2645 & ~42.7 \\
			\bottomrule
		\end{tabular}
	\end{center}
	\label{tbl:Viper}
\end{table}

}

The detailed results of the evaluation are shown separately for each of the files in~\tabref{tbl:Gobra} (Gobra), \tabref{tbl:MPP} (MPP), \tabref{tbl:VerCors} (VerCors), and \tabref{tbl:Viper} (Viper).

\end{document}
\fi

\endinput